\documentclass[manuscript]{aastex631}

\definecolor{Red}{rgb}{1,0,0}

\usepackage{bm}
\usepackage{rotating}
\usepackage{hyperref}

\received{} %June 1, 2019}
\revised{} %January 10, 2019}
\accepted{} %\today}
\submitjournal{ApJ}

\shorttitle{Proton and $\alpha$ Particle Velocity Distributions in Sub-Alfv\'{e}nic Solar Wind}
\shortauthors{Ofman et al.}
\begin{document}

\title{Observations and Modeling of Unstable Proton and $\alpha$ Particle Velocity Distributions in Sub-Alfv\'{e}nic Solar Wind at PSP Perihelia}

\correspondingauthor{Leon Ofman}
\email{ofman@cua.edu}

\author[0000-0003-0602-6693]{Leon Ofman*}
\affiliation{Department of Physics\\
Catholic University of America  \\
Washington, DC 20064, USA}
\affiliation{Heliophysics Science Division\\
NASA Goddard Space Flight Center \\
Greenbelt, MD 20771, USA}
\altaffiliation{Visiting, Department of Geosciences,  Tel Aviv University, Tel Aviv, Israel}

\author[0000-0002-5240-044X]{Scott A Boardsen}
\affiliation{Goddard Planetary Heliophysics Institute\\
 University of Maryland, Baltimore County\\
 Baltimore, MD 21250, USA}
\affiliation{Heliophysics Science Division\\
NASA Goddard Space Flight Center \\
Greenbelt, MD 20771, USA}

\author[0000-0002-6849-5527]{Lan K Jian}
\affiliation{Heliophysics Science Division\\
NASA Goddard Space Flight Center \\
Greenbelt, MD 20771, USA}

\author[0000-0002-3808-3580]{Parisa Mostafavi}
\affiliation{Johns Hopkins University\\
 Applied Physics Laboratory\\
 Laurel, MD 20723, USA}

\author[0000-0001-5030-6030]{Jaye L Verniero}
\affiliation{Heliophysics Science Division\\
NASA Goddard Space Flight Center \\
Greenbelt, MD 20771, USA}

\author{Roberto Livi}
\affiliation{Space Sciences Laboratory\\
University of California\\
Berkeley, CA 94720, USA}

\author{Michael McManus}
\affiliation{Space Sciences Laboratory\\
University of California\\
Berkeley, CA 94720, USA}

\author{Ali Rahmati}
\affiliation{Space Sciences Laboratory\\
University of California\\
Berkeley, CA 94720, USA}

\author[0000-0003-1138-652X]{Davin Larson}
\affiliation{Space Sciences Laboratory\\
University of California\\
Berkeley, CA 94720, USA}

\author[0000-0001-6673-3432]{Michael L Stevens}
\affiliation{Center for Astrophysics Harvard Smithsonian\\
Cambridge, MA 02138, USA}

%\collaboration{1}{(AAS Journals Data Scientists collaboration)}



\begin{abstract}
Past observations show that solar wind (SW) acceleration occurs inside the sub-Alfv\'{e}nic region, reaching the local Alfv\'{e}n speed at typical distances $\sim10-20\ R_s$ (solar radii). Recently,  Parker Solar Probe (PSP) traversed regions of sub-Alfv\'{e}nic SW near perihelia in encounters E8-E12 for the first time providing data in these regions. It became evident that properties of the magnetically dominated SW are considerably different from the super-Alfv\'{e}nic wind. For example, there are changes in relative abundances and drift of $\alpha$ particles with respect to protons, as well as in the magnitude of magnetic fluctuations. We use data of the magnetic field from the FIELDS instrument, and construct ion velocity distribution functions (VDFs) from the sub-Alfv\'{e}nic regions using Solar Probe Analyzer Ions (SPAN-I) data, and run 2.5D and 3D hybrid models of proton-$\alpha$ sub-Alfv\'{e}nic SW plasma. We investigate the nonlinear evolution of the ion kinetic instabilities in several case studies, and quantify the transfer of energy between the protons, $\alpha$ particles, and the kinetic waves. The models provide the 3D ion VDFs at the various stages of the instability evolution in the SW frame. By combining observational analysis with the modeling results, we gain insights on the evolution of the ion instabilities, the heating and the acceleration processes of the sub-Alfv\'{e}nic SW plasma and quantify the exchange of energy between the magnetic and kinetic components. The modeling results suggest that the ion kinetic instabilities are produced locally in the  SW, resulting in anisotropic heating of the ions, as observed by PSP.

\end{abstract}

\keywords{Solar wind (1534); Space probes (1545); Space plasmas (1544); Plasma physics(2089)}


\section{Introduction} \label{intro:sec}

The large-scale, low-frequency solar wind's turbulent fluctuations dissipate at small kinetic scales below the ``break point" scale and this process was recently studied using Parker Solar Probe (PSP) \citep{Fox16} and other spacecraft data \citep[e.g., recently,][]{Bal19,Bow20,Ale21,Tel21}. The heliocentric distance-dependent break point scale occurs at frequencies close to the local proton gyro-resonance frequency \citep[e.g.,][]{Bru13,Tel15}, where the large scale fluctuations in the inertial range provide the energy that eventually cascades via turbulence to the small kinetic scales.  Other plasma processes, such as waves, reconnection, and charged particle beams may dissipate the energy at small scales as well. The resonant conversion of magnetic and kinetic energy to thermal energy of the plasma on ion scales must be facilitated by wave-particle interactions and the corresponding cyclotron resonances and plasma instabilities. Collisions may have a longer timescale long-term and large spatial scale impact on the SW plasma thermalization far from the Sun at $\geq$1AU \citep[e.g.][]{Kas08,Mar13,Tra15,Kas17}. However, in the young solar wind close to the Sun, the role of collisions on the thermalization of non-Maxwellian plasma is negligible and kinetic instabilities may become the dominant thermalization process \citep[e.g.][]{Kas08,Bou13}. Such instabilities have been analyzed using Wind data near 1 AU (215 $R_s$) \citep{Alt18,Alt19,Kas13,Kas17,Mar21} and Helios data at distances as close as 60$R_s$ \citep{Dur19,Dur21}, demonstrating the role of the instabilities in the heating of solar wind plasma. 

Multi-ion plasma instabilities in the SW have been modeled using hybrid-PIC models (using Particle-In-Cell (PIC) model for ions and fluid model for electrons) in the past; for instance, \citet{MVO13} used a 1.5D hybrid model to show how protons and $\alpha$ particles are accelerated by large amplitude Alfv\'{e}n wave spectrum in the expanding SW plasma. \citet{OVM14,OVR17}, \cite{MOV15}, and \citet{OOV15}  used 2.5D hybrid model to study the effects of proton-$\alpha$ super-Alfv\'{e}nic drift in inhomogeneous expanding solar wind plasma as well as evolution of turbulence in solar wind at ion scales. Recently, a number of 3D hybrid models of SW plasma were developed and mainly used to examine the turbulence cascade to ion scales in SW plasma  \citep{Vas15,Vas20,Cer17,Fra18,Hel19,Mar20}. Proton heating by a proton-$\alpha$ drift instability with anisotropic $\alpha$-particle temperature for typical conditions at 1AU was modeled with 1D, 2D, and 3D hybrid models by \citet{Mar22}. A more simplified approach to study the evolution of proton-$\alpha$ anisotropic plasma was developed using macroscopic quasi-linear theory \citep[e.g.,][]{Yoo15}, demonstrating qualitative agreement with PIC models. Recently quasilinear theory was applied to study the proton-$\alpha$ drift instability for a range of typical SW parameters \citep{Sha21}.

Thanks to recent PSP  observations the inner heliosphere, it has become evident that ion kinetic instabilities play an important role in  the dynamics of the SW plasma at an unexpectedly important level in previously unexplored regions close to the Sun \citep[e.g.][]{Vec21,Ver20,Ver22}. The differential streaming of proton-$\alpha$ ion populations was  observed in the young solar wind suggesting preferential ion acceleration processes \citep{Mos22}. Motivated by recent PSP observations at perihelia of proton and $\alpha$ particle beams \citep{Ver20,Ver22}, \citet{Ofm22a} studied the evolution of the beam instabilities in the proton-$\alpha$ super-Alfv\'{e}nic SW plasma using the 2.5D and 3D hybrid codes for conditions close to the Sun. The initial conditions of the models were  setup using the Solar Probe Analyzer - Ions (SPAN-I) \citep{Liv22} data and the associated ion velocity distribution functions (VDFs). It was found that the super-Alfv\'{e}nic ion beams relax in several hundred to several thousand proton gyroperiods, and an  associated spectrum of kinetic ion-scale waves was detected while the ions undergo perpendicular and parallel heating. The strong initial perpendicular heating leads to temperature anisotropy that is subject to the ion-cyclotron instability in accordance with Vlasov theory and hybrid models, followed by nonlinear saturation and relaxation of the unstable distribution through wave-particle interaction.  Evidence of resonant ion-cyclotron wave heating  in observed proton-velocity distributions obtained from SPAN-I was reported recently by \citet{Bow22}. 

In recent perihelia (Encounters 8-12, E8-E12) PSP has entered the magnetically dominated corona \citep{Kas21} and detected consistently the sub-Alfv\'{e}nic solar wind condition, i.e., where the SW speed is below the local Alfv\'{e}n speed, allowing important processes to take place, such as incoming and outgoing Alfv\'{e}n wave interactions associated with turbulent heating. It was found that the sub-Alfv\'{e}nic wind has  different statistical properties of turbulence such as  magnetic fluctuations anisotropy \citep{Ban22}, as well as difference in SW plasma parameters such as $\alpha$ particle abundances and ion temperature anisotropies (see below). The investigation of sub-Alfv\'{e}nic SW conditions is of particular interest for studying SW plasma instabilities, acceleration, and heating mechanisms of the SW plasma. 

In the present study we focus on the sub-Alfv\'{e}nic region observed by PSP during E10-E12, by investigating several cases of SPAN-I and FIELDS observations. We use the proton and $\alpha$ particle VDF data constructed from SPAN-I data in the spacecraft frame to set up 2.5D and 3D hybrid models of the ions in the sub-Alfv\'{e}nic wind, and investigate the nonlinear evolution of the instabilities in the SW frame. The paper is organized as follows. In Section~\ref{obs:sec} we present the PSP data cases and discuss the observational motivation, in Section~\ref{model:sec} we present the hybrid Particle-in-Cell (hybrid-PIC) model details, and in Section~\ref{num:sec} we show the numerical modeling results. The discussion and conclusions are in Section~\ref{disc:sec}.

\section{Observational Motivations} \label{obs:sec}

In this section, we provide the observational motivations for the hybrid modeling study by selecting several cases of interest from PSP data (in the spacecraft frame).  We specifically utilize the data measured by SPAN-I on board the Solar Wind Electron Alpha and Protons investigation (SWEAP) instrument \citep[details in][]{Liv22}. The SPAN-I instrument consists of  a time-of-flight section and an electrostatic analyzer. It is designed to measure the 3D VDF of ions with the energy range of 2 eV to 30 KeV in the solar wind.  The instrument is suited on the ram side of PSP and can only measure the bulk of solar wind when the VDF peak enters the field of view (FOV) of the instrument, which is only a limited time during perihelia. We carefully selected the appropriate time periods for analyzing the VDF observed by PSP.  The  times of interest from different PSP encounters in the sub-Alfv\'{e}nic region during E10-E12 were selected as case studies to initialize the hybrid modeling runs (see below) in order to investigate the onset and nonlinear evolution of ion kinetic instabilities in the solar wind frame.  

\begin{figure}[h]
\centering
\includegraphics[trim={0 2cm 0 0},clip,width=0.77\linewidth]{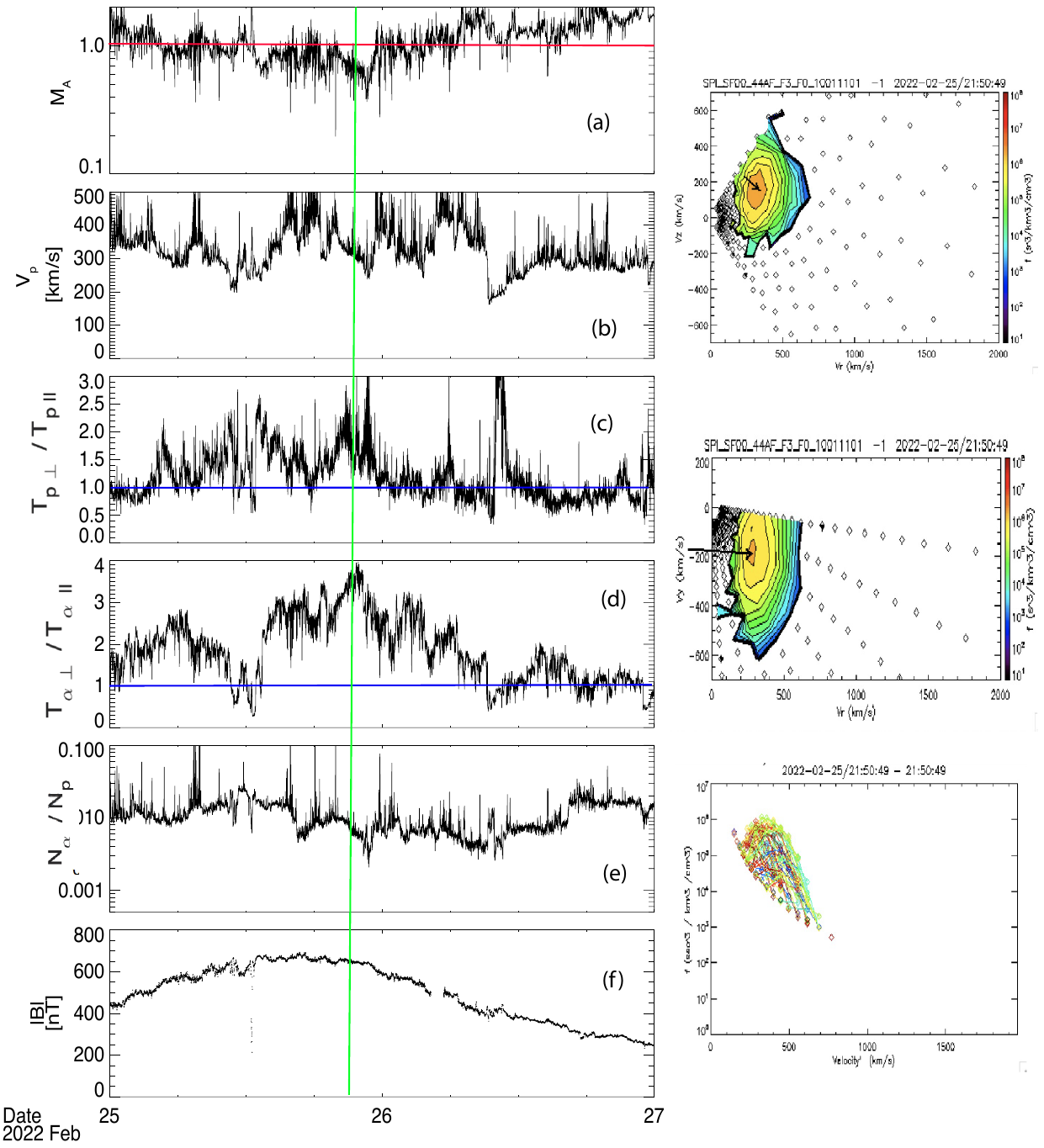}
\caption{An example of the sub-Alfv\'{e}nic region observed by PSP during 2022-02-25 to 2022-02-27 UT during Encounter 11. Left: panels from top to bottom show $M_A$, proton velocity, $T_\perp/T_\parallel$ for proton and $\alpha$ particles, $\alpha$-to-proton density ratio, and $|B|$, respectively. Right panels show the proton VDF at the marked time (green line) on the left panel. The black arrow represents the magnetic field direction in SPAN-I coordinates, where the head is at the solar wind velocity and its length is the Alfv\'{e}n speed. The top right panel shows the proton VDF in the $V_r-V_z$ plane, the middle right panel is the  $V_r-V_y$ plane, and the lower right panel shows lines of energy sweeps at different elevation angles. The obscuring effect of the PSP thermal shield on the observed proton VDF is evident.}
\label{PSPsubalfven02252022E11:fig}
\end{figure}

In Figure~\ref{PSPsubalfven02252022E11:fig}, we show an example of the time interval that includes sub-Alfv\'{e}nic regions, observed by PSP on 2022-02-25 to 2022-02-27 UT during Encounter 11. The left panels from top to bottom show the Alfv\'{e}nic Mach number $M_A$, proton velocity, the  temperature anisotropy $T_\perp/T_\parallel$ for proton and $\alpha$ particles, the $\alpha$-to-proton density ratio, and the magnetic field magnitude $|B|$, respectively. The right panels show the proton VDF constructed from SPAN-I data at the marked time (green line) on the left panel. The black arrow represents the magnetic field direction in SPAN-I coordinates, where the head is at the solar wind velocity and its length is the local Alfv\'{e}n speed. The top right panel shows the proton VDF in the $V_r-V_z$ plane (2D slice through the plane of the elevation angle $\theta$), the middle right panel is the  $V_r-V_y$ plane (2D slice through the plane of the azimuth angle $\phi$), and the lower right panel shows phase-space density of protons for all look directions. 
The green vertical line shows the time on that we focus our analysis. The solar wind is sub-Alfv\'{e}nic (i.e., $M_A<1$ ; Fig. 1a), slow (with the proton velocity $V_p < 400$ km/s;  Fig.~\ref{PSPsubalfven02252022E11:fig}b), large $\alpha$ and proton temperature anisotropies (Figs.~\ref{PSPsubalfven02252022E11:fig}c and \ref{PSPsubalfven02252022E11:fig}d), and small $\alpha$-to-proton density ratio (Fig.~\ref{PSPsubalfven02252022E11:fig}e). 
The right panels show that the selected sub-Alfv\'{e}nic proton VDFs mostly include only a ``core" population, defined as the population centered around the peak in the contour plots. It is also evident that the core is mostly in the field-of-view (FOV) of SPAN-I. 

%1

\begin{figure}[h]
\centering
\includegraphics[trim={0 3cm 0 2cm},clip,width=1.0\linewidth]{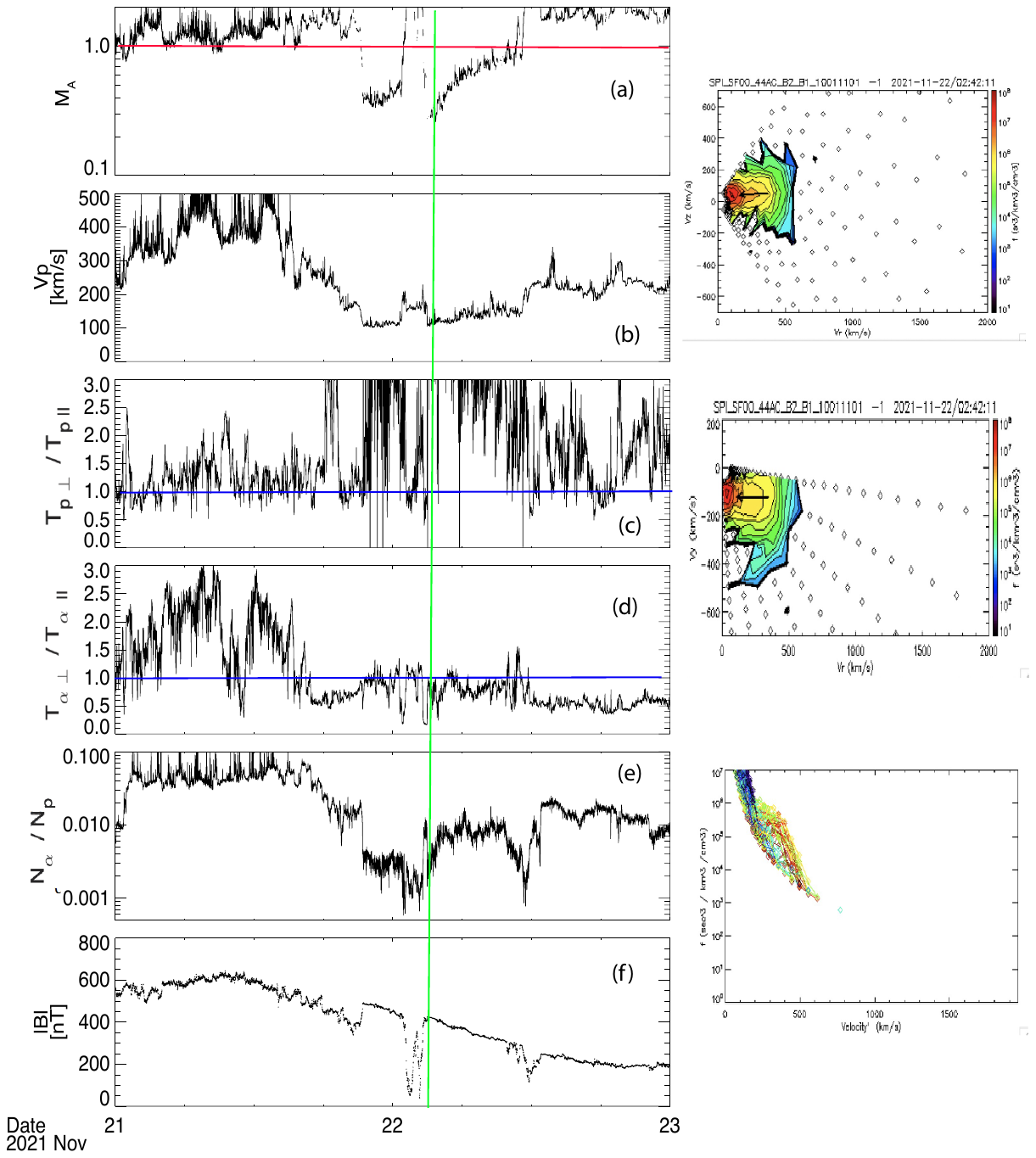}
\caption{An example of the sub-Alfv\'{e}nic region observed by PSP during 2021-11-21 to 2021-11-23  during Encounter 10. The format of the panels is similar to Figure~\ref{PSPsubalfven02252022E11:fig}.}
\label{PSPsubalfven02252022E10:fig}
\end{figure}

In Figure~\ref{PSPsubalfven02252022E10:fig}, another example of sub-Alfv\'{e}nic SW regions is shown as observed by PSP in the time interval 2021-11-21 to 2021-11-23  during Encounter 10. 
%Left: panels from top to bottom show Alfv\'{e}nic Mach number, $M_A$, proton velocity, temperature anisotropy $T_\perp/T_\parallel$ for proton and $\alpha$ particles, $\alpha$-to-proton density ratio, and the magnetic field magnitude $|B|$, respectively. Right panels show the proton VDF constructed from SPAN-I data at the indicated times as the marked with the green line on the left panel. The black arrow represents the magnetic field direction in SPAN-I coordinates, where the head is at the solar wind velocity and its length is the Alfv\'{e}n speed. The top right panel shows the proton VDF in the $V_r-V_z$ plane, the middle right panel is the  $V_r-V_y$ plane, and the lower right panel shows lines of  energy sweeps at different elevation angles. The obscuring effect of the PSP thermal shield on the observed proton VDF is evident.
The panels in this figure are similar to the ones in Figure~\ref{PSPsubalfven02252022E11:fig}. Note that here the solar wind is very slow ($V_p\sim$ 100 km s$^{-1}$, see, Fig.~\ref{PSPsubalfven02252022E10:fig}b) and the $\alpha$ temperature anisotropy is small (Fig.~\ref{PSPsubalfven02252022E10:fig}d). The obscuring effect of the PSP thermal shield on some parts of the observed proton VDF is evident. The panels on the right show mostly one or two proton populations (core and beam) with evidence of a third population in some of the cases. 
%2

\begin{figure}[h]
\centering
\includegraphics[width=\linewidth]{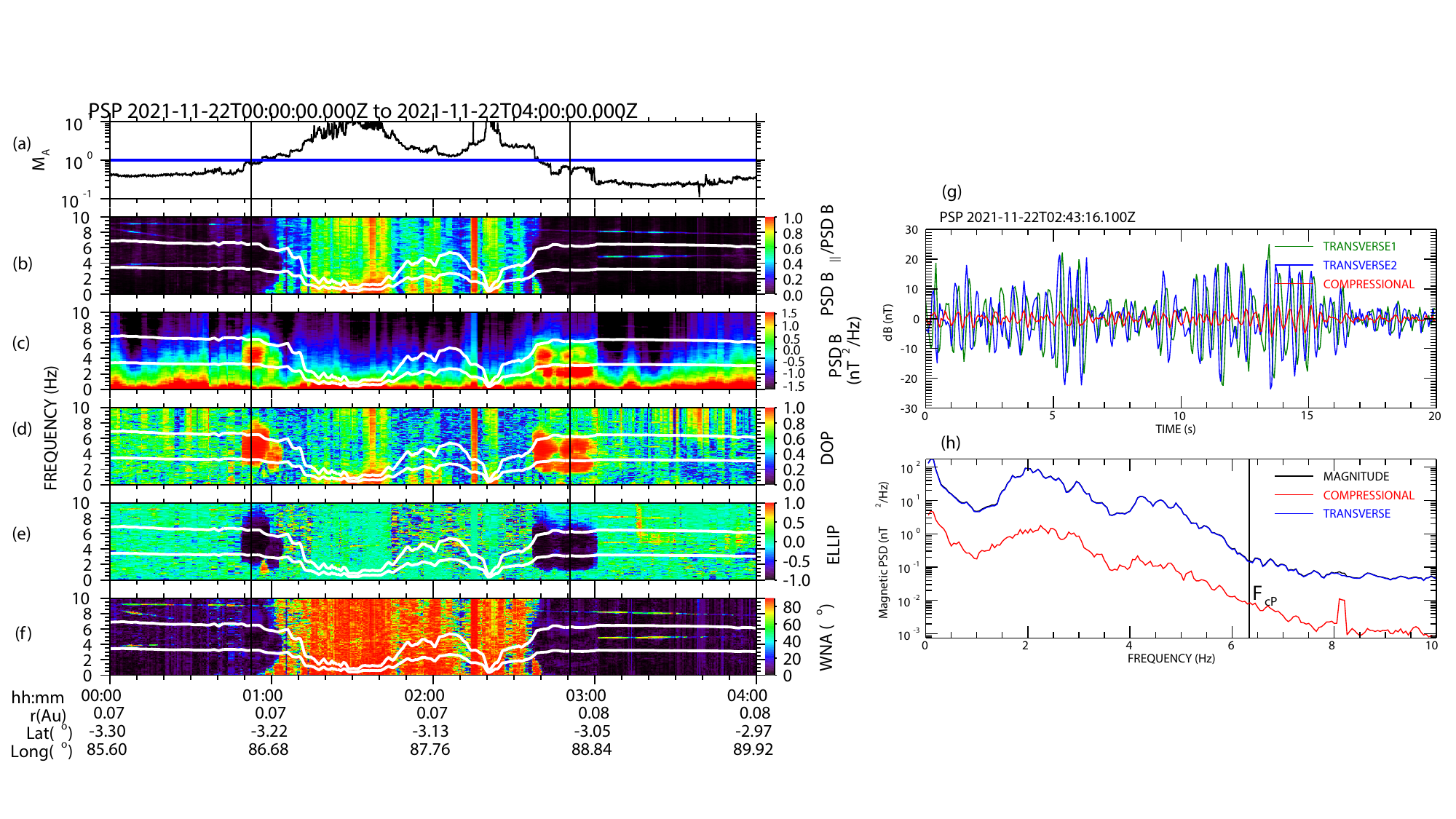}
\caption{An example of kinetic wave activity analysis of the magnetic field observed by PSP FIELDS instrument during 2021-11-22 00:00UT to 04:00UT. Left panels show (a) $M_A$, (b) magnetic compressibility $(\delta B_\parallel/\delta B)^2$, (c) magnetic power spectral density, (d) degree of polarization (DOP), (e) ellipticity, and (f) wave normal angle (WNA) in degrees. The two white lines (b-f) are at the proton and $\alpha$ cyclotron frequencies. Right panels show the (g) linearly detrended  magnetic fluctuations $\delta B$  time series in field aligned coordinates for 20 s  and the (h) power spectrum with the compressional (red) and transverse (blue) polarization. The proton cyclotron frequency, $F_{cp}$, is indicated by the vertical line.}
\label{PSPsubalfven11212021E10waves:fig}
\end{figure}

%Figure~\ref{PSPsubalfven11212021E10waves:fig}, shows an example of kinetic wave activity analysis of the magnetic field observed by PSP FIELDS instrument during 2021-11-21 18:00UT to 2021-11-22 04:00UT. Left panels show $M_A$, power spectral density (PSD) of  $B_\parallel$/PSD $B$, degree of polarization (DOP), ellipticity, and wave normal angle (WNA) in degrees, respectively. Right panels show the magnetic fluctuations $\delta B$ (top panel) for the time interval 02:43:16 to 0258:40UT on 2021-11-22, and the wave power spectrum (lower panel) with the compressional (red) and transverse (blue) polarization. The proton gyroresonat frequency, $F_{cp}$, is indicated by the vertical line. 
Figure~\ref{PSPsubalfven11212021E10waves:fig} shows an example of ion-scale wave activity analysis based on the flux gate magnetometer (FGM instrument  of PSP FIELDS suite) measurements of the magnetic field observed during the time interval 2021-11-22 00:00 UT to 04:00 UT. Left panels show (a) $M_A$, (b) magnetic compressibility $(\delta B_\parallel/\delta B)^2$, (c) magnetic power spectral density (PSDB), (d) degree of polarization (DOP), (e) ellipticity, and (f) the wave normal angle (WNA). The two white lines in panels (b-f) are at the proton and $\alpha$ particle cyclotron frequencies.  The FGM data sampled at $\sim$293 samples/s was down sampled to 20 samples/s so the Nyquist frequency is 10 Hz. The polarization analysis was computed on the down sampled data using the first method described by \citet{Art76} from which the degree of polarization, ellipticity, and the wave normal angle were computed. The spectral matrix used in the polarization analysis was computed from Fast Fourier Transforms (FFTs) of the magnetic field components with a window size of 20 s, staggered in time by 20 s and the spectral matrix was averaged over the nearest neighbors in both time and frequency. Define $<B>$ as the average of the magnetic field over each 20 s interval. The degree of polarization is a function of the coherency of the spectral components in the plane normal to the wave vector direction: a value near 0 indicates no coherence, while near 1 indicates strong coherence. The ellipticity is measured in the plane perpendicular to $<\mbox{\bf B}>$, for circularly polarized waves its sign is -1 (1) if the transverse magnetic field fluctuation $\delta \mbox{\bf B}_\perp$ rotates in the left (right) handed sense about $<\mbox{\bf B}>$. The wave normal angle (folded into the 1st quadrant) is the angle between the wave vector and $<\mbox{\bf B}>$. The magnetic compressibility is the ratio of the magnetic parallel power spectral density divided by the total magnetic power spectral density. In the sub-Alfv\'{e}nic interval, the wave power in this frequency range is observed to be primarily transverse. During the super Alfv\'{e}nic part of the interval, the compressibility is large, but this is likely due to strong step-like transitions in the magnetic field magnitude and is possibly an artifact of the FFT. Examining the polarization panels (c-e) in Figure~\ref{PSPsubalfven11212021E10waves:fig}, two regions characterized by peaks in PSD, of DOP $\sim 1$ and ellipticity $\sim$ -1 are observed: one centered at $\sim$00:55 UT and the other centered at $\sim$2:50 UT, $M_A$ $\sim$1 for the former and $M_A$ $\sim$0.7 for the later. Within the measurement error, the wave normal angle is at 0$^\circ$ during these two intervals.

%3

\begin{figure}[h]
\centering
\includegraphics[width=\linewidth]{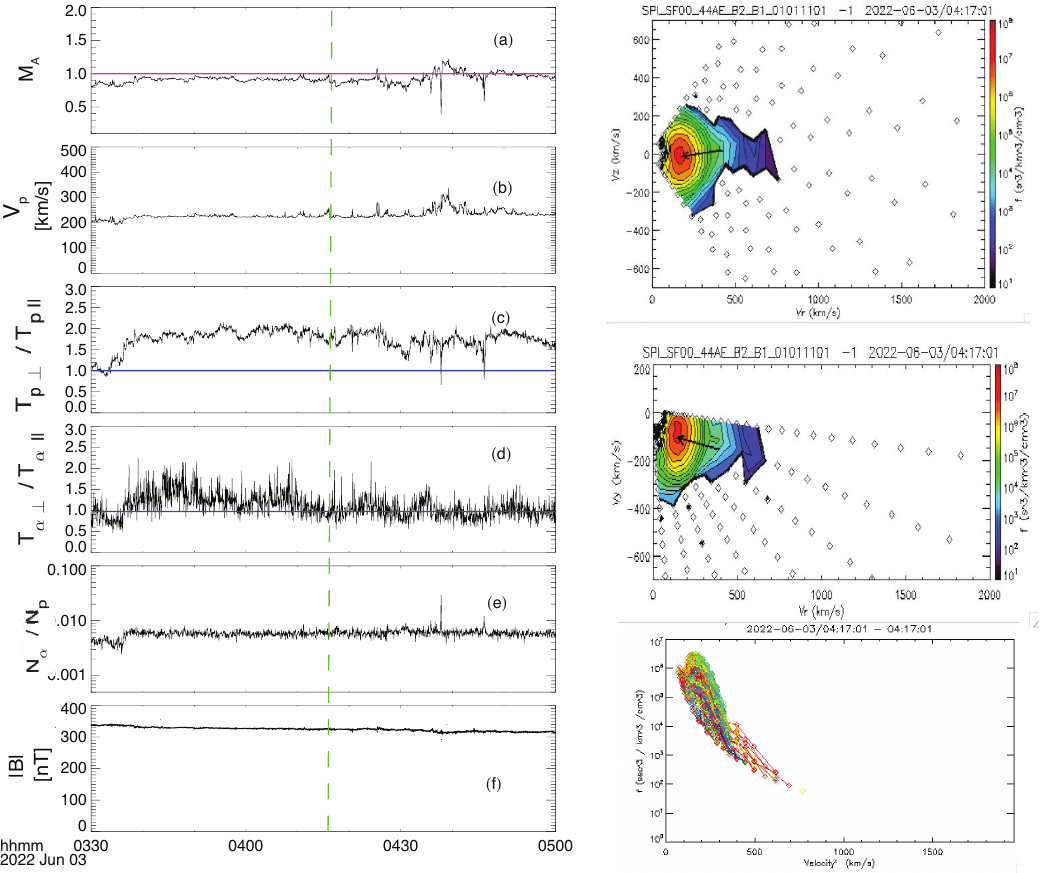}
\caption{An example of the sub-Alfv\'{e}nic region observed by PSP on 2022-06-03 03:30-5:00 UT during Encounter 12. The format of the panels are similar to Figure 1.}
\label{PSPsubalfven06032022E12:fig}
\end{figure}

In Figure~\ref{PSPsubalfven06032022E12:fig}, we show the analyzed PSP data during a few hours of Encounter 12, examining the plasma properties and VDFs inside the sub-Alfv\'{e}nic region. The value of the Alfv\'{e}nic Mach number shows that the solar wind is magnetically dominated (Figure~\ref{PSPsubalfven06032022E12:fig}a). The panels in this figure are similar to the ones in Figure~\ref{PSPsubalfven02252022E11:fig}. The solar wind properties such as proton velocity, proton and $\alpha$ particles temperature anisotropies, and $\alpha$-to-proton density ratio are plotted in Figure~\ref{PSPsubalfven06032022E12:fig}b-e. The magnetic field magnitude during this time interval is plotted in the lower panel. The right panels in Figure~\ref{PSPsubalfven06032022E12:fig} show the VDF of protons at the marked time on the left panel by green dashed line. It clearly shows the presence of temperature anisotropy and a super-Alfv\'{e}nic proton beam separated from its core. However, it is evident that the relative number of protons in the beam population is extremely small (note the log scale of the VDF). Thus, we focus on the proton core temperature anisotropy as the main source of the kinetic instability and associated ion-scale wave generation, while the beam has a negligible contribution in this particular case. 

%4

In Table~\ref{pspdat:tab}, we summarize the observational parameters of sub-Alfv\'{e}nic solar wind obtained with PSP FIELDS and SPAN-I instruments at E10-E12 in several cases for our study, shown in Figures~\ref{PSPsubalfven02252022E11:fig}-\ref{PSPsubalfven06032022E12:fig} at the given dates/times.  The proton velocity, $V_p$,  the $\alpha$ particle velocity $V_\alpha$, and the Alfv\'{e}n speed $V_A$ are given in km/s.  The proton and $\alpha$ particle number densities $n_{p}$, $n_\alpha$ are per $cm^{-3}$. The dimensionless Alfv\'{e}nic Mach number $M_A$ is given in each case. The temperatures in parallel $T_{\parallel}$ and perpendicular $T_{\perp}$ directions with respect to the magnetic field direction are given in eV. The corresponding temperature anisotropies $A_p\equiv T_{\perp,p}/T_{\parallel,p}$ for protons and  $A_\alpha\equiv T_{\perp,\alpha}/T_{\parallel,\alpha}$ for the $\alpha$ particles are given in the table. The magnetic field magnitude $|B|$ is in nT and the corresponding proton gyrofreqency is in the range of $5\sim10$ Hz. The Mach number, the Alfv\'{e}n speed, $V_A$, and the $p-\alpha$ drift speed, $V_d$, are computed using the observed magnetic field, densities and velocities. We compute the Alfv\'en speed as $V_A = B/\sqrt{\mu_0 (n_p m_p + n_{\alpha} m_{\alpha})}$, where B is the magnetic field magnitude, $\mu_0$ is the permeability of vacuum, and $n_{p/\alpha}$ and $m_{p/\alpha}$ the proton/$\alpha$ number density, and their mass, respectively. We also calculate the $\alpha$-proton differential drift speed using 
$V_{d} = sign(|V_{\alpha}| - |V_p|) |V_{\alpha} - V_p|$ \citep{Ste96,Dur17,Mos22}.
%(Steinberg et al. 1996; Durovcov́a et al. 2017; Mostafavi et al. 2022). 
This equation considers the directions of both vectors (see more details in \citet{Mos22}). 
These data products are publicly available online at NASA Space Physics Data Facility (SPDF) \href{https://spdf.gsfc.nasa.gov/}{https://spdf.gsfc.nasa.gov/}. In the following sections, we study and compare the properties of the solar wind protons and $\alpha$ particles, as well as associated kinetic-scale wave activity inside the sub-Alfv\'{e}nic solar wind regions to understand the kinetic evolution and the heating of the ions.

\section{Hybrid-PIC Model, Initial and Boundary Conditions} \label{model:sec}
In order to model the $p-\alpha$, magnetized SW plasma, we employ our recently developed parallelized 2.5D (i.e., two spatial dimensions and three components of the velocities and fields) and 3D hybrid codes, which are based on the same principles as the 1D hybrid code developed by \citet{WO93}. The model uses the particle-in-cell (PIC) method for the ions, while the electrons are modeled as a background neutralizing massless fluid, i.e., by solving the generalized Ohm's law. The code was expanded to 2D \citep{OV07}, parallelized \citep{Ofm10}, and expanded to a full 3D model \citep{Ofm19b}. In the present study, we use primarily the 2.5D hybrid  with a supporting computation with full 3D hybrid modeling.  In the 2.5D model, the currents and fields are calculated on a 256$^2$ grid  with up to 512 particles per cell (ppc), as in our previous studies of ion kinetic instabilities in the SW \citep{OVM14,OVR17,Ofm22a}. The required limitation on the overall statistical noise and the VDF resolution, which is typically a function of $\beta$, determine the required number of particles per cell. A numerical convergence test and total energy conservation monitoring are typically used to determine that the noise level is below the physical fluctuations level in the hybrid codes. We note that a Gaussian charge distribution occupies more than a cell and represents many physical particles  used to model each numerical super-particle, reducing statistical noise and hence the required number of numerical particles compared to more simplified numerical models. The ion kinetic dissipation scale determines the shortest resolution length scales in warm multi-ion plasma. We have found strong damping of the resonant wave branches at $k_\parallel \delta>0.5$, where the normalization is by the inverse proton inertial length $\delta=c/\omega_{pp}$, where $\omega_{pp}=\sqrt{4\pi n_pe^2/m_p}$ is the proton plasma frequency and $\Omega_p=eB/(m_pc)$ is the proton gyrofrequency that is used to normalized the time. The solution of Vlasov’s linear dispersion relation as well as previous hybrid models reveal that the proton cyclotron waves are significantly dampened for values of $k_\parallel\delta\gtrsim 1$ \citep[e.g.,][]{OV07,OVM14,OVR17}. As a result, the length scales that correspond to the kinetic dissipation range and lower are well resolved by our model with resolution of $0.75\delta$, and the largest scale is more than an order of magnitude higher than the dissipation scale.  We use the typical time step $\Delta t\sim0.025\Omega_p^{-1}$, so that the proton gyroperiod is well resolved. 

The equations of motions of each ion are solved subject to the electromagnetic field forces of each cell in the 2D or 3D domain, while the currents and charges are calculated by summing over the particle charges and velocities, that are used to update the fields in each cell. The boundary conditions are periodic and the equations are solved using the 2$^{\rm nd}$ order Rational Runge-Kutta (RRK) method \citep{Wam78}. Further  details of the model equations can be found in \citet{OVM11,Ofm19b}. In the present model we do not consider the effects of the solar wind expansion, due to the short duration of the relaxation of the instabilities of several hundred gyroperiods compared to the typical expansion time of $10^6 \Omega_p^{-1}$ for the present SW parameters. The initial VDFs of the protons and $\alpha$ particles are setup with bi-Maxwellian or drifting Maxwellian distributions with the parameters such as $T_{i,\perp}$ and $T_{i,\parallel}$, $n_i$, $\beta_i$ (where $i=p,\ \alpha$),  the $\alpha$-proton drift speed, $V_d$, for the various cases adopted from the PSP data (see Table~\ref{pspdat:tab}). The VDFs quickly become non-Maxwellian as a result of the evolution of the instabilities, as demonstrated below. For simplicity we obtain $n_e$ from the quasi-neutrality condition, assume $\beta_e=\beta_p$ with isothermal electron pressure in the model, while it was shown in past modeling studies that the details of the electron thermal structure have small effects on the ion kinetic instability results. The velocities in the model are normalized with the Alfv\'{e}n speed, $V_A$ in each case. When there is strong evidence of a beam (such as in Case~3) the bi-Maxwellian initial state does not capture well the observation based VDF and it is preferable to construct a core-beam distribution as in our recent study \citep{Ofm22a}. In section~\ref{num:sec} we present the numerical results for several cases of interest.

\section{Linear Stability Analysis of the Initial State}

The stability threshold for temperature anisotropy driven instabilities for protons as well as for heavier ions, such as the $\alpha$ particles in the $\beta_{\parallel,i}-T_\perp/T_\parallel$ parameter space, was derived in parametric form using linear Vlasov's theory for protons  \citep[e.g.,][]{Gar92,Gar01a}. This expression for protons applicable in the range  $0.05<\beta_{\parallel,p}<5$ is given by 
\begin{eqnarray}
&&\frac{T_{\perp,p}}{T_{\parallel,p}}-1=\frac{S_p}{\beta_{\parallel,p}\,^{a_p}},
\label{Apbetap:eq}
\end{eqnarray} where the parameters $S_p\sim 1$, and $a_p\approx 0.4$ independent of the growth rate.  The relation was generalized for heavier ions and tested using 1.5D and 2.5D hybrid-PIC models \citep{Gar01b,OVG01}, given by
\begin{eqnarray}
&&\frac{T_{\perp,i}}{T_{\parallel,i}}-1=\frac{S_i}{[(m_p/m_i)\tilde{\beta}_{\parallel,i}]\,^{a_i}},
\label{Aibetai:eq}
\end{eqnarray} where $\tilde{\beta}_{\parallel,i}=8\pi n_e k_B T_i/B^2$, where $n_e$ is the electron number density, and the parameters $S_i\gtrsim 1$, and $a_i\approx 0.4$ independent of the growth rate. In the present study $i=\alpha$ and  $m_p/m_i=1/4$. Thus, using the parametric relation, Equations~\ref{Apbetap:eq}-\ref{Aibetai:eq}, we estimate the marginal instability condition and find that the proton population in Case~1 in Table~\ref{pspdat:tab} with  $A_p=4.3$, $\beta_{\parallel,p}=0.08$ is expected to be unstable with respect to the ion-cyclotron instability, while Cases~2 and 4 are closely below the linear instability threshold for protons. Nevertheless, Cases~2 and 4 are found to be unstable with the nonlinear hybrid model (see, Section~\ref{num:sec} below). A similar expression to Equation~\ref{Apbetap:eq} for the resonant firehose instability when $T_{\perp,p}/T_{\parallel,p}<1$ \citep{Gar98} shows that in Case~3 the protons are below the linear instability threshold, and is also stable in nonlinear hybrid model results below.  The electron number density, $n_e$, can be approximated as $n_p+2n_\alpha$ using the quasi-neutrality condition, and neglecting minor ions in the SW. In all the considered cases in Table~\ref{pspdat:tab} the $\alpha$ particle population is below the instability threshold, the proton-$\alpha$ drift is below the Alfv\'{e}n speed, and hence does not significantly affect the stability. However, we note that the linear instability thresholds are approximate, and in some linearly marginally stable cases the SW plasma may  exhibit kinetic instability with slow growth rate when the fully nonlinear hybrid model is applied.
 
\section{Numerical Results} \label{num:sec}

In this section we present the numerical results of the 2.5D and 3D hybrid models for the cases initialized with the observed PSP parameters and show the results of the models in Figures~\ref{vdfpE11Ap4.3Aa1.6vd0.3:fig}-\ref{VDFE11p_3d:fig}.

\subsection{2.5D Hybrid Modeling Results}

\begin{figure}[ht]
\centering
\includegraphics[trim={0 0 0 3cm},clip,width=\linewidth]{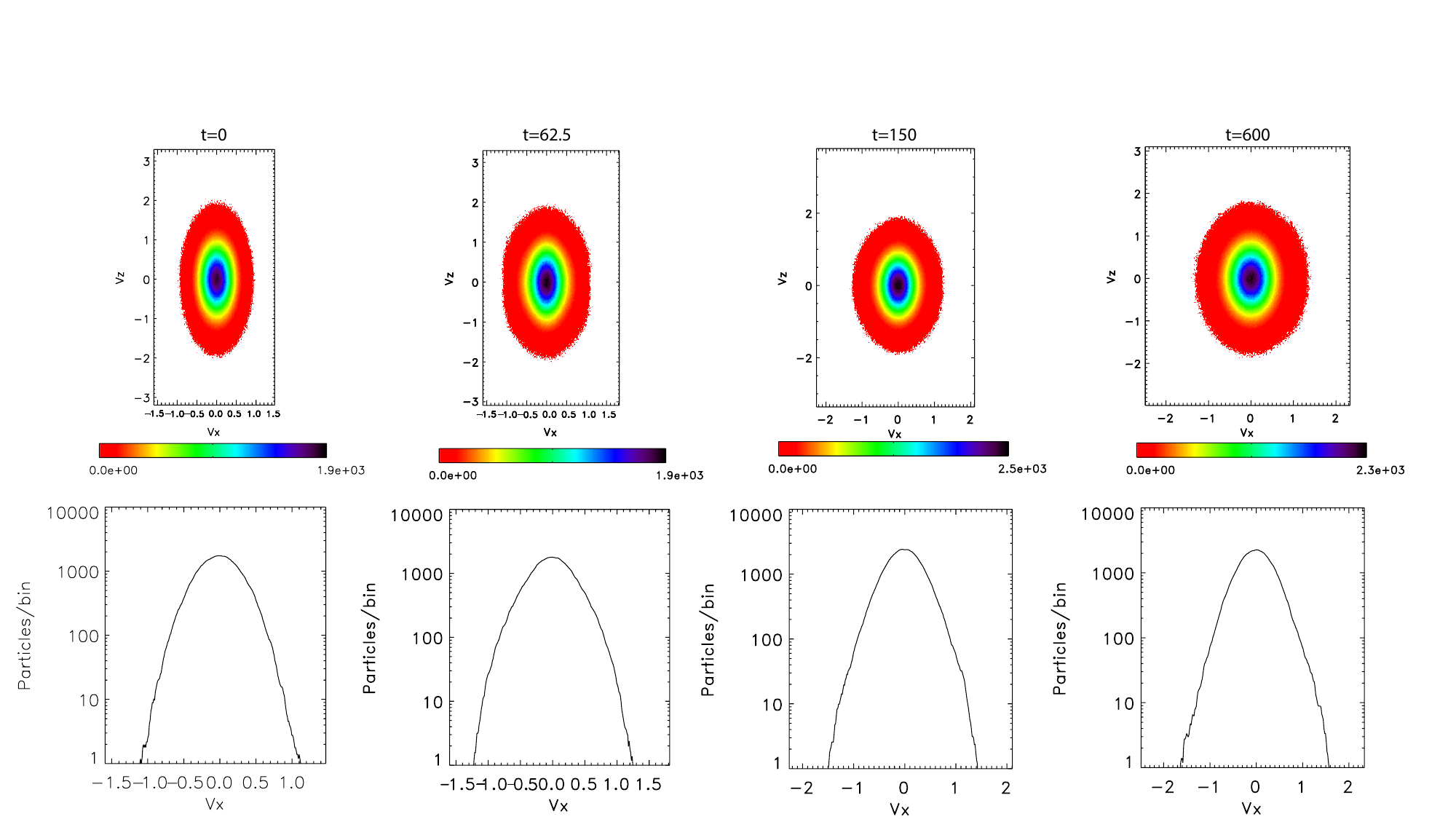}
\caption{ The temporal evolution of the proton VDF for Case~1 with the initial parameters of E11 at 2022-02-25 21:50:50 UT
with $A_p=4.3$, $A_\alpha=1.6$, $n_\alpha/n_p=0.007$, and $V_d=0.32V_A$. The upper panels show the proton VDFs in $V_x-V_z$ plane, and the lower panels show the cut along $V_x$ through the corresponding peaks of the proton VDFs at $t=0,\ 62.5,\ 150,\ 600\Omega_p^{-1}$.}
\label{vdfpE11Ap4.3Aa1.6vd0.3:fig}
\end{figure}

The results of the 2.5D hybrid model for Case~1 with the initial parameters of E11 at 2022-02-25 21:50:50 UT
with $A_p=4.3$, $A_\alpha=1.6$, $n_\alpha/n_p=0.007$, and $V_d=0.32$ are shown in Figures~\ref{vdfpE11Ap4.3Aa1.6vd0.3:fig}-\ref{anisoE11Ap4.3Aa1.6vd0.3:fig}. In Figure~\ref{vdfpE11Ap4.3Aa1.6vd0.3:fig} we show the initial state and the temporal evolution of the proton VDF in $V_x-V_z$ phase space plane, and the cuts along $V_x$ through the corresponding peaks at times $t=0,\ 62.5,\ 150,\ 600\Omega_p^{-1}$. It is evident that the initially unstable anisotropic state with $A_p=4.3$ relaxes towards more isotropic VDF at the final state ($t=600\Omega_p^{-1}$) with $A_p=2$. It is also evident that the  shape of the distribution remains elongated in the perpendicular direction of the velocity space throughout the evolution, suggesting that the departure from the bi-Maxwellian VDF is small for the protons in this case, and the cut along $V_x$ exhibits a Maxwellian shape throughout the evolution. 
%5
\subsubsection{Velocity Distributions}
\begin{figure}[h]
\centering
\includegraphics[trim={0 0 0 2.5cm},clip,width=\linewidth]{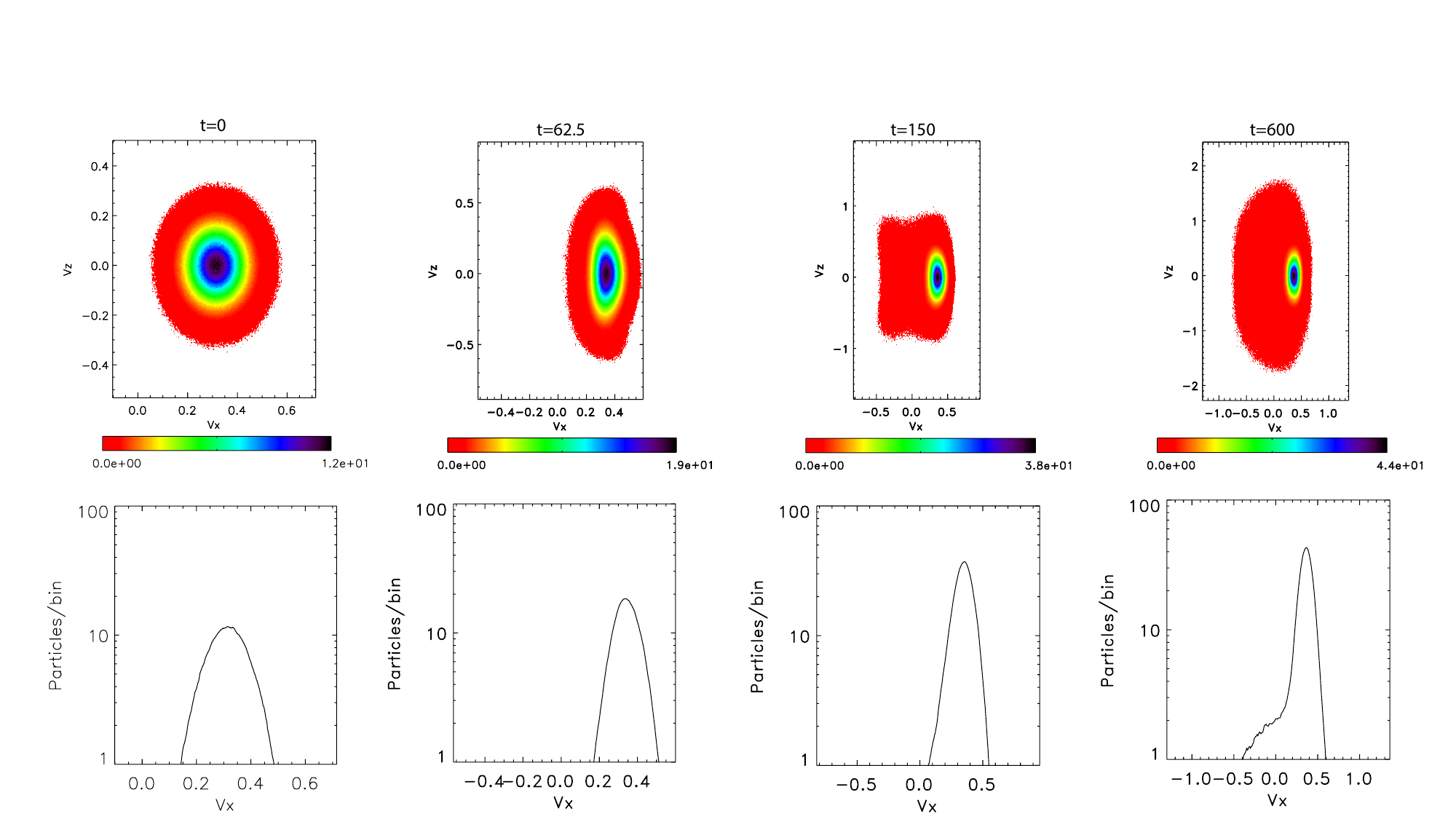}
\caption{ The temporal evolution of the $\alpha$ particle VDF for Case~1 with the initial parameters of E11 at 2022-02-25 21:50:50 UT
with $A_p=4.3$, $A_\alpha=1.6$, $n_\alpha/n_p=0.007$, and $V_d=0.32V_A$. The upper panels show the $\alpha$ particle VDFs in $V_x-V_z$ plane, and the lower panels show the cut along $V_x$ through the corresponding peaks of the $\alpha$ particle VDFs at $t=0,\ 62.5,\ 150,\ 600\Omega_p^{-1}$.}
\label{vdfaE11Ap4.3Aa1.6vd0.3:fig}
\end{figure}

In Figure~\ref{vdfaE11Ap4.3Aa1.6vd0.3:fig} the initial VDF in the $V_x-V_z$ plane of the $\alpha$ population and at $t=62.5,\ 150,\ 600\Omega_p^{-1}$ are shown. The initial state of the $\alpha$ particles VDF is anisotropic drifting bi-Maxwellian with $A_\alpha=1.6$ and the drift velocity $V_d=0.32V_A$. However, very quickly at $t=62.5\Omega_p^{-1}$ the anisotropy of the $\alpha$ particle VDF increases due to the resonant perpendicular heating by the proton-emitted spectrum and become non-gyrotropic. The perpendicular heating of the $\alpha$ particle population is relatively strong `per-particle' due to the small relative to protons abundance of $\alpha$ particles resonating with the electromagnetic ion cyclotron (EMIC) wave spectrum emitted by the proton-cyclotron instability. The Doppler shift due to the drift velocity affects the resonant condition $\omega-k_\parallel v_\parallel=\pm\Omega_i$, where $i=p, \alpha$,  and $v_\parallel\approx V_d$, increasing the net perpendicular heating. This has been verified by repeating the run with $V_d=0$, that showed significantly smaller $\alpha$ particle heating. At $t=150\Omega_p^{-1}$ the $\alpha$ particles start develop a beam structure, with stronger evident beam at  $t=600\Omega_p^{-1}$. The beam formation proceeds due to the wave-particle interactions with the proton-emitted kinetic waves spectrum that scatter the $\alpha$ particles towards the proton population velocity space peak.
%6
\subsubsection{Temporal evolution of anisotropies, energies, and drift speed}
\begin{figure}[h]
\centering
\includegraphics[trim={0 0 0 2cm},clip,width=\linewidth]{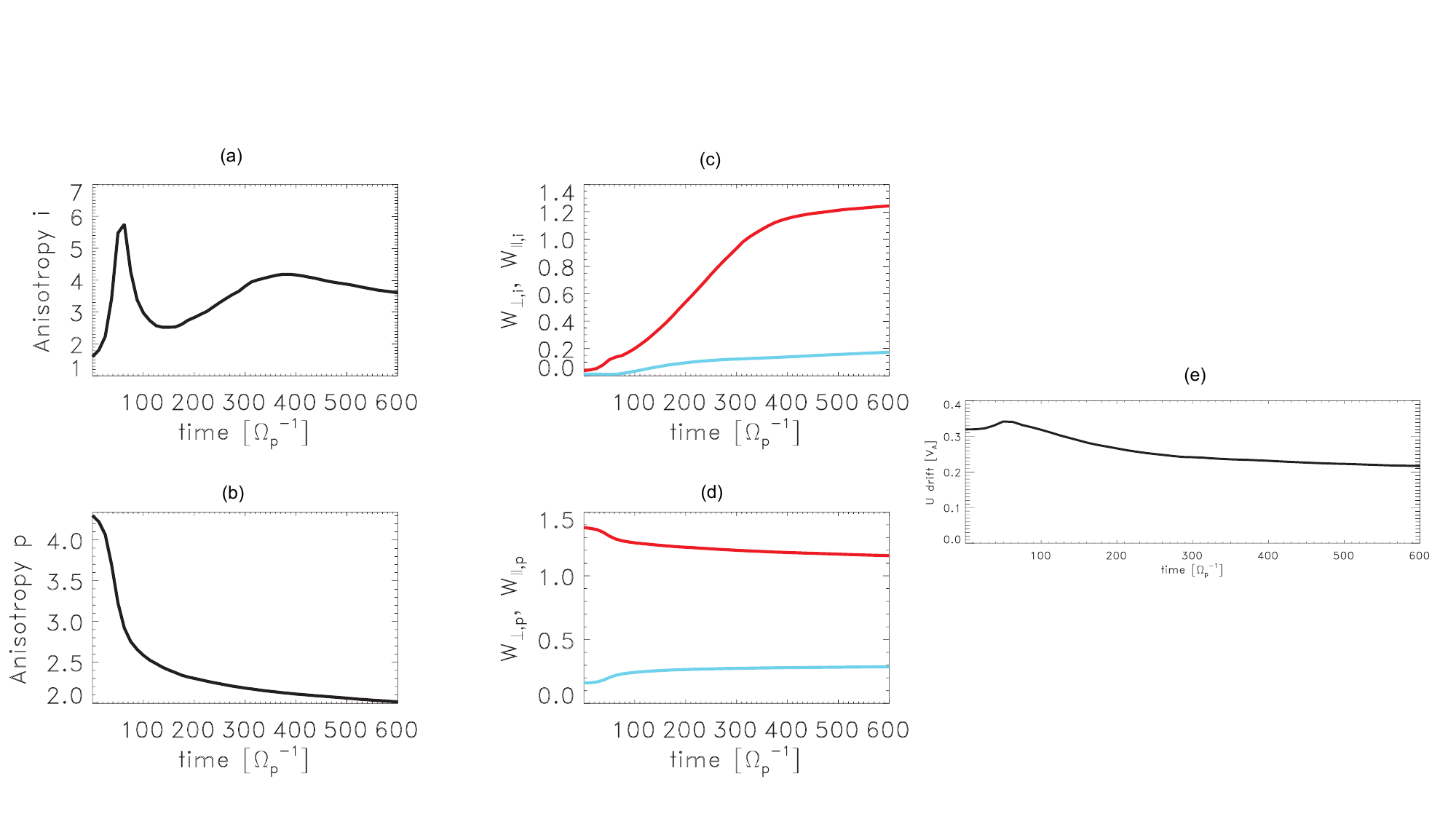}
\caption{ The temporal evolution for Case~1  with the initial parameters of E11 at 2022-02-25 21:50:50 UT with $A_p=4.3$, $A_\alpha=1.6$, $n_\alpha/n_p=0.007$, and $V_d=0.3V_A$ of (a) the $\alpha$ particle temperature anisotropy, (b) proton temperature anisotropy, (c) $\alpha$ particles perpendicular kinetic energy, $W_{\perp,i}$, (red), parallel kinetic energy, $W_{\parallel,i}$, (blue), (d) proton perpendicular kinetic energy, $W_{\perp,p}$, (red), parallel kinetic energy, $W_{\parallel,p}$, (blue), and (e) the drift velocity $V_d$.}
\label{anisoE11Ap4.3Aa1.6vd0.3:fig}
\end{figure}

Figure~\ref{anisoE11Ap4.3Aa1.6vd0.3:fig} is devoted to the temporal evolution of the temperature anisotropies, parallel, $W_{\parallel, i}$ and perpendicular, $W_{\perp, i}$, proton and $\alpha$ kinetic energies, and the proton-$\alpha$ drift velocity for Case~1 with the initial parameters $A_p=4.3$, $A_\alpha=1.6$, $n_\alpha/n_p=0.007$, and $V_d=0.3V_A$ are shown for times $t=0-600\Omega_p^{-1}$. Figure~\ref{anisoE11Ap4.3Aa1.6vd0.3:fig}a shows the $\alpha$ particle temperature anisotropy that increases rapidly from an initial value of 1.6 to close to 6 in $\sim65\Omega_p^{-1}$. The rapid increase is due to resonant perpendicular heating of the minor ion population, where the $\alpha$ particles resonate with part of the ion-cyclotron spectrum emitted by the relaxation of the unstable proton VDF. The increased $\alpha$ particle temperature anisotropy is highly unstable, and relaxes rapidly through the secondary ion-cyclotron instability of the $\alpha$ population, with evidence of additional tertiary weaker instability at later time, and final anisotropy at $t=600\Omega_p^{-1}$ of about 3.6, more than double the initial $A_\alpha$. The rapid relaxation of the proton temperature anisotropy is evident in Figure~\ref{anisoE11Ap4.3Aa1.6vd0.3:fig}b on a timescale of $\sim80\Omega_p^{-1}$, similar to the rapid perpendicular heating timescale of the $\alpha$ particle population. The proton population temperature anisotropy relaxes to 2.0 at the end of the run ($t=600\Omega_p^{-1}$). The strong perpendicular heating of the $\alpha$ particles is also evident in the temporal evolution of kinetic energy in Figure~\ref{anisoE11Ap4.3Aa1.6vd0.3:fig}c, where the $W_{\perp,\alpha}$ increases much faster and reaches higher final value compared to $W_{\parallel,\alpha}$. During this time period, the proton population exhibits perpendicular cooling, where $W_{\perp, p}$ decreases by $\sim20\%$, and parallel heating, where $W_{\parallel, p}$ increases by about  a factor of two thanks to the relaxation of the ion-cyclotron instability and wave particle scattering that results in velocity space diffusion from the perpendicular to parallel direction (see Figure~\ref{anisoE11Ap4.3Aa1.6vd0.3:fig}d). The initial sub-Alfv\'{e}nic proton-$\alpha$ drift of 0.32$V_A$ is stable with respect to the magnetosonic drift instability where the instability threshold is $\gtrsim V_A$ \citep[see, e.g.][]{Gar93,XOV04,OV07,Ver13}. It is evident in Figure~\ref{anisoE11Ap4.3Aa1.6vd0.3:fig}e that initially $V_d$ is increasing by $\sim$8\%, peaking at the time of maximal $\alpha$ heating, following by gradual decrease to 0.22$V_A$ at the end of the run at $t=600\Omega_p^{-1}$. While the initial $p-\alpha$ population drift has no significant effect on the protons due to the small relative abundance of the $\alpha$ particles, the drift does affect the perpendicular heating of the $\alpha$ particle population. By repeating the model run with parameters the same as in Case~1, but with  $V_d=0$, it becomes evident that the drift increases the initial anisotropy of the $\alpha$ population since the  the Doppler shift caused by the drift velocity affects the resonance of the $\alpha$ with the proton-emitted ion cyclotron wave spectrum. While there is some qualitative agreement in the evolution of the temperature anisotropies and drift velocity in Case~1 with the quasilinear results of \citet{Sha21} in their Case~4, the model parameters are substantially different for direct comparison.

%7

\begin{figure}[h]
\centering
\includegraphics[trim={0 0 0 2cm},clip,width=\linewidth]{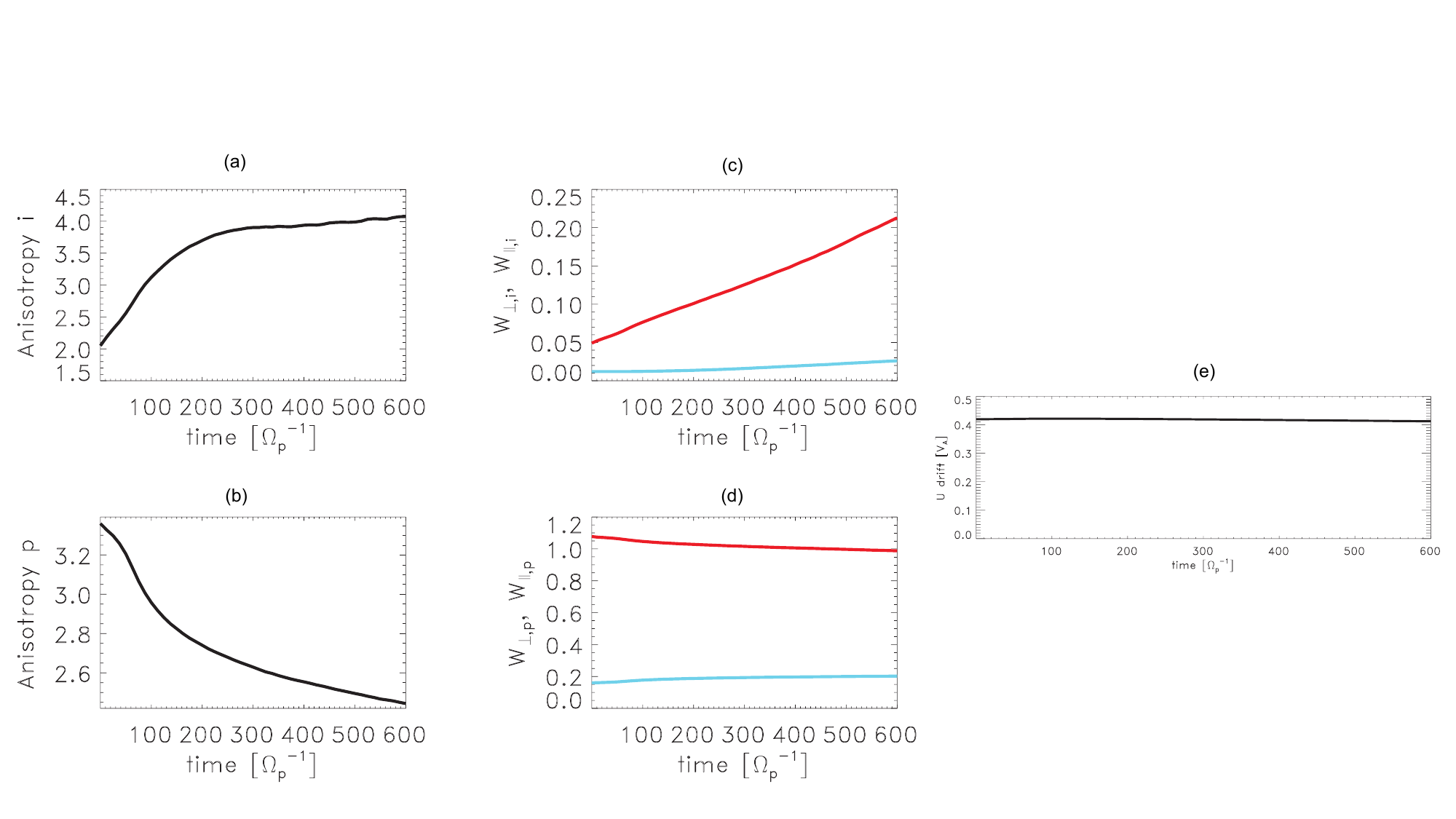}
\caption{The temporal evolution for Case~2 with the initial parameters of E11 at 2022-02-25 18:56:38 UT with initial $A_p=3.3$, $A_\alpha=2$, $n_\alpha/n_p=0.007$, and $V_d=0.42V_A$ of (a) the $\alpha$ particle temperature anisotropy, (b) proton temperature anisotropy, (c) $\alpha$ particles perpendicular kinetic energy, $W_{\perp,i}$, (red), parallel kinetic energy, $W_{\parallel,i}$, (blue), (d) proton perpendicular kinetic energy, $W_{\perp,p}$, (red), parallel kinetic energy, $W_{\parallel,p}$, (blue), and (e) the drift velocity $V_d$.}
\label{anisoE11Ap3.3Aa2vd0.4:fig}
\end{figure}

In Figure~\ref{anisoE11Ap3.3Aa2vd0.4:fig} we show the temporal evolution of the 2.5D hybrid model for Case~2 of E11 at 2022-02-25 18:56:38 UT with the initial parameters $A_p=3.3$, $A_\alpha=2$, $n_\alpha/n_p=0.007$, and $V_d=0.42V_A$. The evolution of the ion temperature anisotropies, perpendicular and parallel energies, and the $p-\alpha$ drift speed are shown. It is evident that relaxation of the proton anisotropy in Case~2 is slower than in Case~1 since the initial anisotropy $A_p=3.3$ is lower than in the previous case, relaxing to $\sim2.4$ at the end of the run  (see Figure~\ref{anisoE11Ap3.3Aa2vd0.4:fig}b). The corresponding time $\alpha$ particles temperature anisotropy increases  more gradually in Case~2  compared to Case~1, reaching final anisotropy of $\sim 4.1$ at $t=600\Omega_p^{-1}$. The evolution of the $\alpha$ particle perpendicular $W_{\perp,i}$ and parallel $W_{\parallel,i}$ kinetic energies is shown in Figure~\ref{anisoE11Ap3.3Aa2vd0.4:fig}c-d. It is evident that the $\alpha$ particle population undergoes strong perpendicular heating, while the protons are cooling in the perpendicular direction. Both protons and $\alpha$ particles are heated gradually in the parallel direction. The initial $p-\alpha$ relative drift, $V_d=0.42$, is below the threshold of the magnetosonic drift instability, and the drift speed remains nearly constant throughout the temporal evolution (Figure~\ref{anisoE11Ap3.3Aa2vd0.4:fig}e). 
%8

\begin{figure}[ht]
\centering
\includegraphics[width=\linewidth]{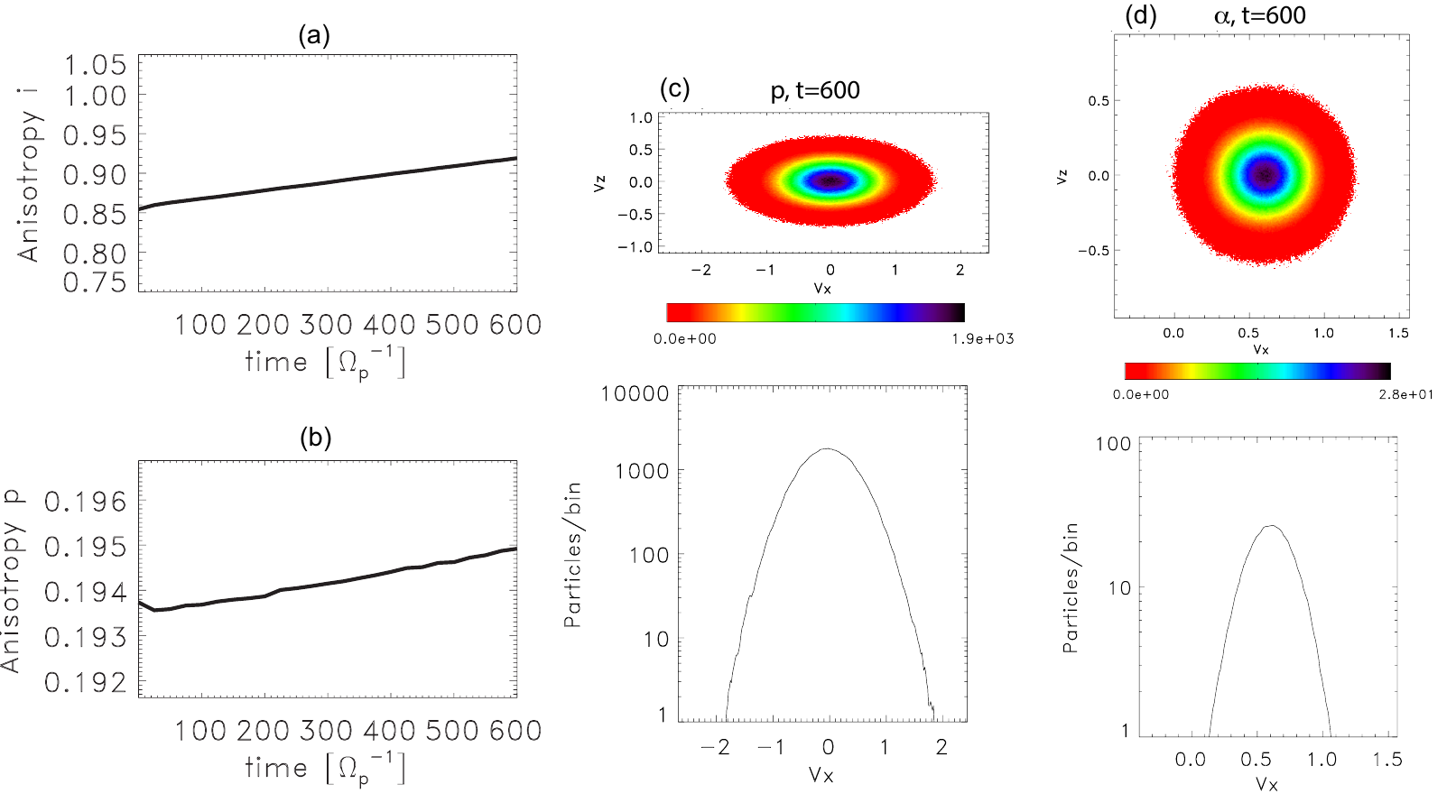}
\caption{ The temporal evolution of the proton VDF for Case~3 with the initial parameters of E10 at 2021-11-22 02:42:10 UT
with $A_p=0.19$, $A_\alpha=0.85$, $n_\alpha/n_p=0.013$, and $V_d=0.62V_A$. (a) The temporal evolution of the $\alpha$ particle temperature anisotropy (note the small range of the $y$-axis). (b) The temporal evolution of the proton temperature anisotropy (note the small range of the $y$-axis). (c) The proton VDFs in $V_x-V_z$ plane at $t=600\Omega_p^{-1}$ (top) the cut along $V_x$ through the peak of the proton VDF (lower panel). (d) The $\alpha$ particle VDFs in $V_x-V_z$ plane at $t=600\Omega_p^{-1}$ (top) the cut along $V_x$ through the peak of the $\alpha$ particle VDF (lower panel).}
\label{anisovdfpE10Ap0.2Aa0.85vd0.62:fig}
\end{figure}

We have performed a 2.5D hybrid model run using the initial parameters of Case~3 with $A_p=T_{p,\perp}/T_{p,\parallel}=0.19$, $A_\alpha=T_{\alpha,\perp}/T_{\alpha,\parallel}=0.85$ that were used to initialize bi-Maxwellian VDFs for the ions, and with $n_\alpha/n_p=0.013$, $V_d=0.62V_A$. The main results for Case~3 are shown in Figure~\ref{anisovdfpE10Ap0.2Aa0.85vd0.62:fig}. The temporal evolution of the $\alpha$ population anisotropy shows slight growth form the initial $A_\alpha=0.85$ towards isotropization with $A_\alpha=0.92$ at the end of the run (Figure~\ref{anisovdfpE10Ap0.2Aa0.85vd0.62:fig}a). The proton anisotropy remains practically unchanged throughout the run to  $t=600\Omega_p^{-1}$ within about 1\%  (Figure~\ref{anisovdfpE10Ap0.2Aa0.85vd0.62:fig}b). The initially bi-Maxwellian proton VDF remains nearly unchanged at the end of the run, with the final VDF in $V_x-V_z$ plane and the $V_x$ direction shown in Figure~\ref{anisovdfpE10Ap0.2Aa0.85vd0.62:fig}c, with very small changes. Similarly, the $\alpha$ particle VDF exhibits small change throughout the modeling run with the final VDF shown in Figure~\ref{anisovdfpE10Ap0.2Aa0.85vd0.62:fig}d. We have also found that the proton-$\alpha$ drift speed remains nearly constant throughout the modeling run. Thus, the modeling results show that Case~3 does not exhibit any numerically detectable ion kinetic instability, in agreement with linear Vlasov stability analysis. For example, where for $A_p=0.2$ the condition $\beta_p\sim1$ must be satisfied for significant growth rate of the firehose instability \citep[see, e.g.,][pp. 129-130]{Gar93}, while in the present Case~3 we have $\beta_p=0.1$ from PSP data. We note that in the present case the bi-Maxwellian initial state constructed from the temperature anisotropy of the protons does not represent well the VDF obtained from SPAN-I data, shown in the right panels of Figure~\ref{PSPsubalfven02252022E10:fig}. The non-Maxwellian proton VDF  shows evidence of proton beam-core state with an Alfv\'{e}nic beam, as discussed and modeled in \citet{Ofm22a}.

%9

\begin{figure}[ht]
\centering
\includegraphics[width=0.5\linewidth]{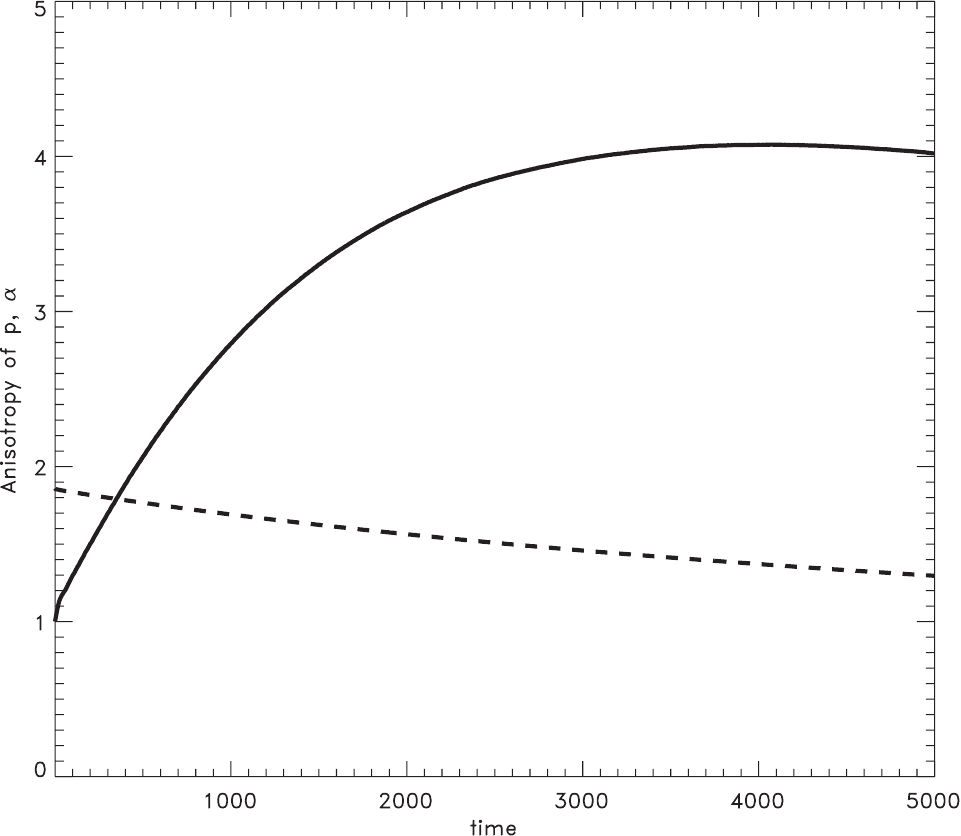}
\caption{ The temporal evolution of the temperature anisotropy of $\alpha$ particles (solid) and protons (dashes) for Case~4 with the initial parameters of E12 at 2022-06-03 04:17:01 UT with $A_p=1.86$, $A_\alpha=1.0$, $n_\alpha/n_p=0.007$, $V_d=0.32V_A$.}
\label{anisoE12_long:fig}
\end{figure}

We have investigated the growth and relaxation of ion-cyclotron instability in the sub-Alfv\'{e}nic SW in Case~4, with the parameters obtained during E12 on 2022-06-03/04:17:01UT, where the initial $A_p=1.86$, $\beta_p=0.1$, and the $\alpha$ particles were initially isotropic. The relative proton-$\alpha$ drift was $V_d=0.32V_A$ and did not change throughout the modeling run. We found that the protons population is unstable with respect to the ion-cyclotron instability, with small growth rate $<10^{-3}\Omega_p$, where the initial proton anisotropy relaxes to an isotropic VDF over $t>5\times10^3\Omega_p^{-1}$, while the $\alpha$ particle population is heated gradually in the perpendicular direction to $A_\alpha\approx4$, reaching maximal anisotropy in $t\sim4\times10^3\Omega_p^{-1}$, following by gradual decrease (see Figure~\ref{anisoE12_long:fig}). Since the evolution is extremely long, the modeling run was not extended to follow the expected gradual relaxation of the $\alpha$ particle temperature anisotropy toward an equilibrium state.

%10

\subsubsection{Wave dispersion and magnetic energy evolution}
\begin{figure}[h]
%\centering
%\hspace{-2cm}\includegraphics[width=0.7\linewidth]{dispE11Ap4.3Aa1.6vd0.3.pdf}
%\hspace{-2cm}\includegraphics[width=0.7\linewidth]{dispE11Ap3.3Aa2vd0.4.pdf}
\includegraphics[width=0.66\linewidth]{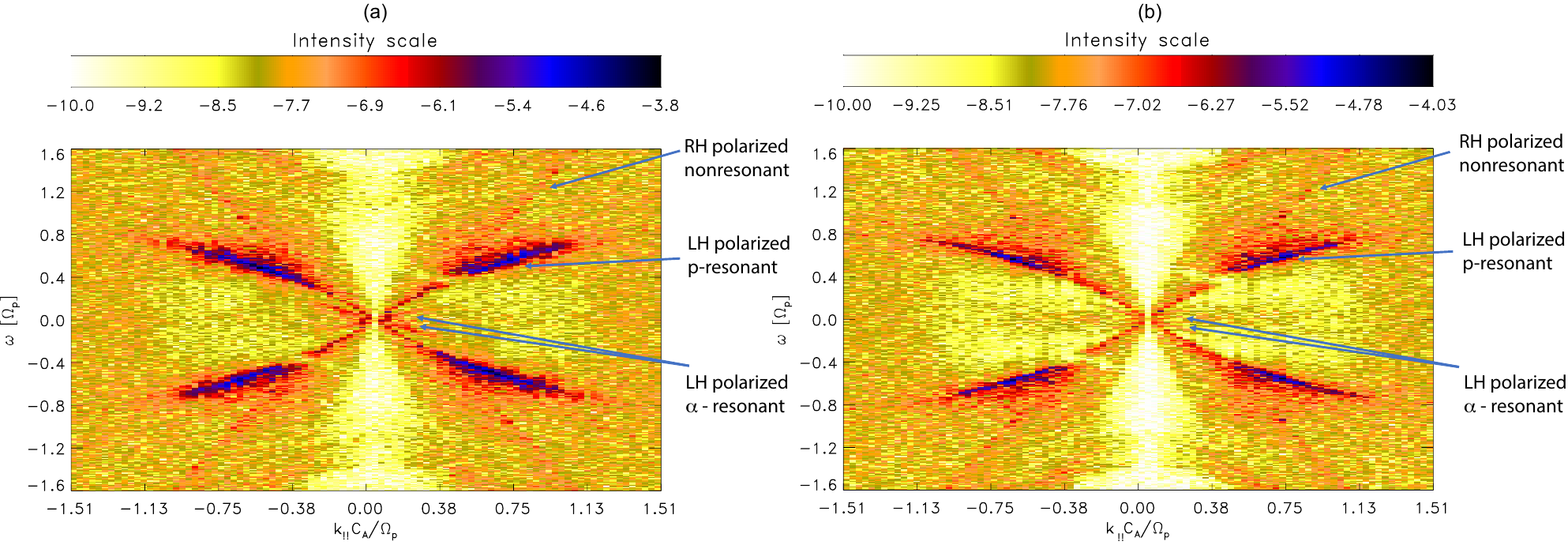}
\includegraphics[width=0.33\linewidth]{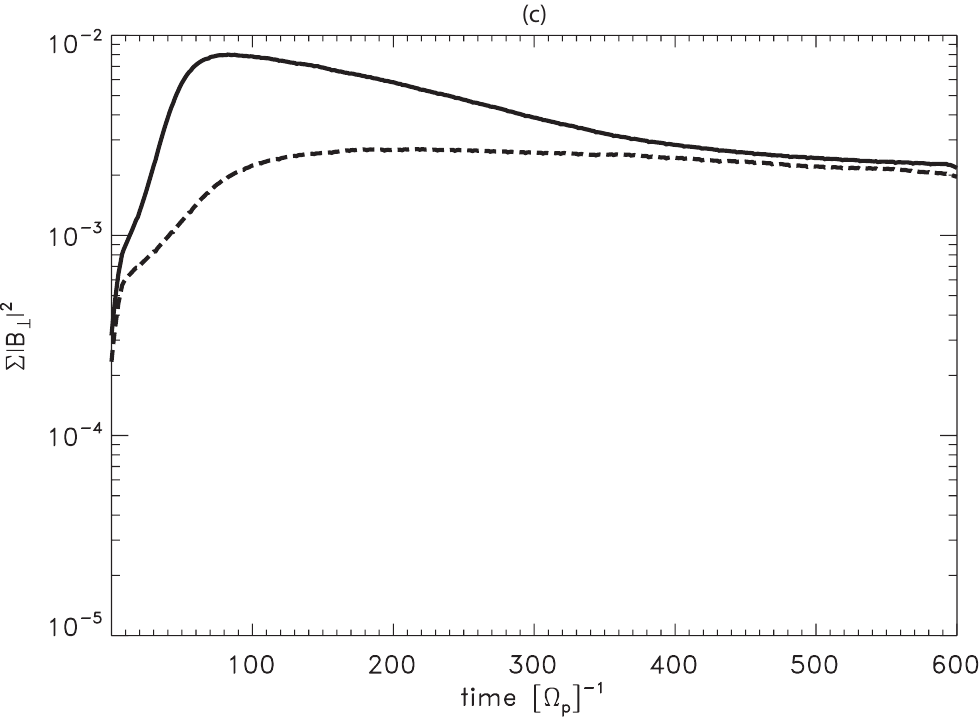}
\caption{ The dispersion relation obtained from the 2.5D hybrid modeling results for (a) Case~1 with the initial parameters of E11 at 2022-02-25 21:50:50 UT with $A_p=4.3$, $A_\alpha=1.6$, $n_\alpha/n_p=0.007$, and $V_d=0.3V_A$. The LH resonant proton and $\alpha$  as well as the nonresonant RH polarized dispersion branches are marked with arrows and labeled. (b) Case~2 on 2022-02-25 at 18:56:38 with the initial parameters  $A_p=3.4$, $A_\alpha=2.0$. The power is shown on a $log_{10}$ intensity scale, indicating that the dominant power is in the proton resonant branches. (c) The modeled total perpendicular magnetic energy for Case~1 (solid) and Case~2 (dashes).}
\label{dispE11fig}
\end{figure}

The dispersion relation computed from the 2.5D hybrid modeling results for Cases 1 and 2 are shown in Figure~\ref{dispE11fig}. The dispersion was constructed by Fourier transforming the spatial and the temporal fluctuations of the $B_\perp$ magnetic field component. In Figure~\ref{dispE11fig}a the dispersion relation of Case~1 is shown with the initial parameters of E11 at 2022-02-25 21:50:50 UT with $A_p=4.3$, $A_\alpha=1.6$, $n_\alpha/n_p=0.007$, and $V_d=0.3V_A$. The left-hand (LH) polarized resonant proton branches contain the most wave power peaking at normalized $|k_\parallel|\approx0.7$, and the $\alpha$  LH resonant branches are evident (marked with arrows). In Figure~\ref{dispE11fig}b the dispersion relation of Case~2 is shown for E11 on 2022-02-25 at 18:56:38 UT with the initial parameters  $A_p=3.4$, $A_\alpha=2.0$. It is evident that the less unstable case shows lower power in the proton resonant branches, and the $\alpha$ population branches are weak, even though their initial temperature anisotropy is larger than in Case~1. This is due to the fact that the $\alpha$ population is heated primarily by the proton instability generated kinetic wave spectrum. It is evident that the proton resonant branches are damped for $|k_\parallel|\gtrsim1$, while the $\alpha$ population resonant branches are strongly damped for $|k_\parallel|\gtrsim 0.3$. The non-resonant (with the ions) right-hand (RH) polarized fast mode branches are evident in Figure~\ref{dispE11fig}a, b, and indicated by the arrows. These branches are of similar power in Cases~1 and 2, and proceed to increase with $|\omega|$ as $|k_\parallel|$ is increased.  The structure of the computed dispersion relation branches is in good agreement with the linear solution of Vlasov dispersion relation for warm proton-$\alpha$ plasma with relative proton-$\alpha$ drift  \citep[e.g.][]{XOV04,OV07,MOV15}. The dispersion relation computed from the hybrid model using the fully nonlinear solution provides additional information of the power, and the damping of the waves. The modeled total perpendicular magnetic energy for Case~1 (solid) and Case~2 (dashes) is shown in Figure~\ref{dispE11fig}c. It is evident that the initial total perpendicular magnetic energy, $\Sigma |B_\perp|^2$, in the more unstable Case~1 reaches about four times larger value in $t\approx70\Omega_p^{-1}$, compared to the peak in Case~2. However, in both cases the wave energy eventually reduces to similar asymptotic values at the end of the runs, as the magnetic energy is converted to kinetic (thermal) energy of the ions in a gradual evolution.

%11

\subsection{3D Hybrid Modeling Results}

\begin{figure}[h]
\centering
\includegraphics[trim={0 0 0 3cm},clip,width=\linewidth]{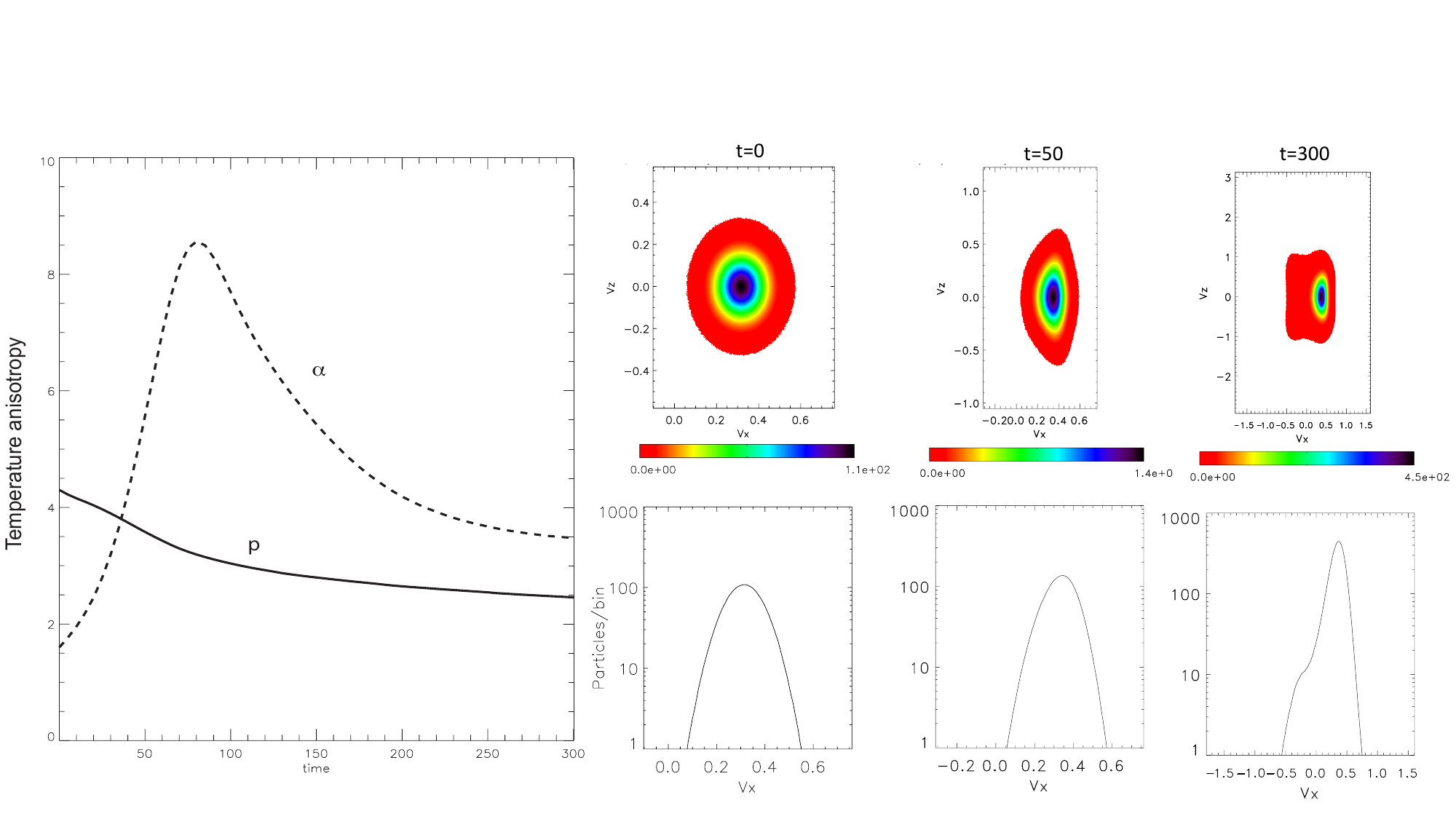}
\caption{The results of the 3D hybrid modeling run with the parameters of Case~1 of E11. The left panel shows the temporal evolution of the temperature anisotropy for protons (solid), and $\alpha$ particle (dashes). The right panels show the $\alpha$ particle VDFs at $t=0,\ 50,\ 300\Omega_p^{-1}$ in the $V_x-V_z$ plane (upper panels) and the cuts of the VDF along $V_x$ (lower panels).}
\label{anisoVDFE11a_3d:fig}
\end{figure}

\begin{figure}[h]
\centering
\includegraphics[trim={0 0 0 3cm},clip,width=\linewidth]{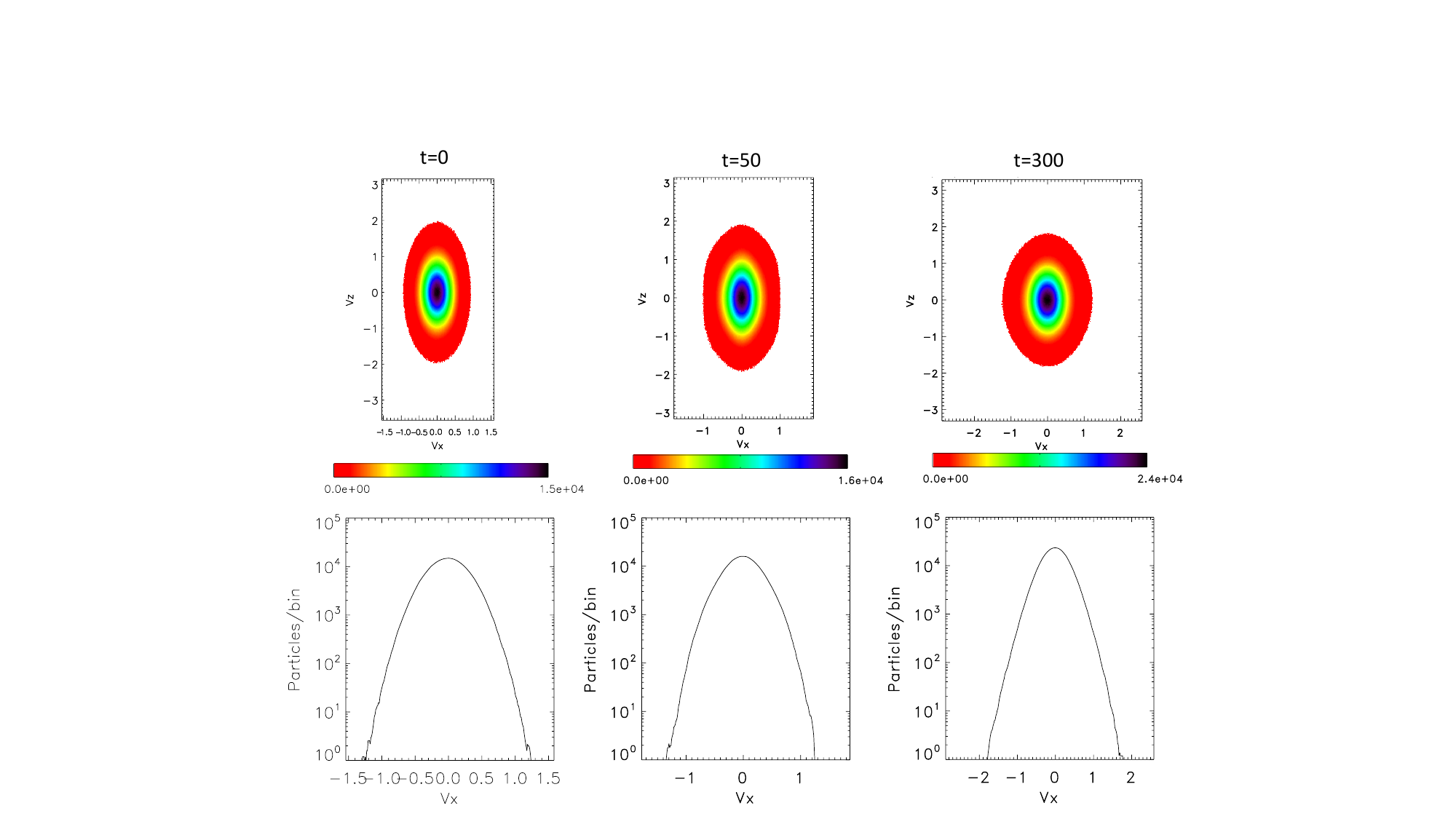}
\caption{The results of the 3D hybrid modeling run with the parameters of Case~1 of E11. The panels show the proton VDFs in the $V_x-V_z$ plane (upper panels) and the cuts of the VDFs along $V_x$ (lower panels) at $t=0,\ 50,\ 300\Omega_p^{-1}$.}
\label{VDFE11p_3d:fig}
\end{figure}

In order to validate the 2.5D hybrid modeling results with more realistic 3D hybrid model, we repeat the run  with the parameters of Case~1 for $t=300\Omega_p^{-1}$ using the computationally intensive 3D hybrid model and present the results in Figures~\ref{anisoVDFE11a_3d:fig}-\ref{VDFE11p_3d:fig}. In Figure~\ref{anisoVDFE11a_3d:fig} (left panel) we show the temporal evolution of the proton and $\alpha$ particle temperature anisotropies. The initially ion-cyclotron unstable anisotropic proton distribution relaxes rapidly from an anisotropy of $A_p=4.3$ at $t=0$ to 2.3 at $t=300\Omega_p^{-1}$. This evolution is very close to the 2.5D hybrid modeling results, where the temperature anisotropy relaxes to 2.2 in the same time interval. At the same time, the $\alpha$ particles are heated in the perpendicular direction, reaching a peak anisotropy of $A_\alpha\approx8.5$ at $t\approx80\Omega_p^{-1}$ with the final $\alpha$  anisotropy of 3.5 at the end of the run at $t=300\Omega_p^{-1}$. This evolution is similar to the the 2.5D hybrid modeling result, where the $\alpha$ anisotropy peaks at $\sim$6 in $\sim65\Omega_p^{-1}$ and is 3.7 at $t=300\Omega_p^{-1}$. Thus, the main difference between the 2.5D and 3D hybrid models are in the details of the peak $\alpha$ particle perpendicular heating, while the proton evolution is close in both models. 

 In Figures~\ref{anisoVDFE11a_3d:fig} (upper right panels) we show the $\alpha$ particle VDF in the $V_x-V_z$ plane and the cuts of the VDFs along $V_x$ (lower right panels) at the initial state, $t=0$, at $t=50$ where the perpendicular heating of the $\alpha$ population are in the most rapid stage, and at the end of the run at $t=300\Omega_p^{-1}$ where the $\alpha$ particles are in the most relaxed stage in this model run. The increased anisotropy is most evident in the elongated shape of the VDF in the $V_z$ direction and the $\alpha$ drift velocity  is evident in the location of the peak VDF in $V_x$ in the upper and lower panels. The final stage of the evolution shows the formation of the $\alpha$ particle beam in the VDF at $t=300\Omega_p^{-1}$, where some of the $\alpha$ particles diffuse in velocity space due to the wave-particle scattering towards the peak of the proton distribution. The evolution of the proton VDF in the $V_x-V_z$ plane and the cuts of the VDFs along $V_x$ are shown in Figure~\ref{VDFE11p_3d:fig}. The proton VDF at $t=0$ exhibits the initial bi-Maxwellian structure with $A_p=4.3$, and the relaxation of the unstable VDF proceeds at $t=50\Omega_p^{-1}$ with evident slight departure from elliptical  bi-Maxwellian shape. However, at the end of the run at $t=300\Omega_p^{-1}$ the proton VDF regains the bi-Maxwellian shape with lower temperature anisotropy. The evolution of both, $\alpha$ particle and proton VDFs, is close to the results obtained with the 2.5D hybrid model, shown in Figures~\ref{vdfpE11Ap4.3Aa1.6vd0.3:fig}-\ref{vdfaE11Ap4.3Aa1.6vd0.3:fig} above.

%12

%13

%\begin{table}[h]
\begin{sidewaystable}[ht]
\scriptsize
%\footnotesize

\caption{The observationally derived parameters of sub-Alfv\'{e}nic solar wind for several time intervals obtained from PSP data at E10-E12. }
%\begin{center}
\centering
%\vspace{-0.1in}
%\begin{tabular}{ | r | r | r | r |}
\begin{tabular}{llcccccccccccccccccc}
\hline
\# &Date/time                & Enc. & $V_p$  & $V_\alpha$  & $n_{p}$  & $n_\alpha$ &  $n_{\alpha}/n_e$ & $\beta_p$ & $\beta_{\alpha}$ & $M_A$ & $V_A$ & $V_d$ & $T_{p,\parallel}$ & $T_{p,\perp}$ & $A_p$ & $T_{\alpha,\parallel}$ & $T_{\alpha,\perp}$ & $A_\alpha$ & $|B|$ \\
  &  &  & [km/s] & [km/s] & [cm$^{-3}$] & [cm$^{-3}$] &  & & & & [km/s] & [$V_A$]& [eV] & [eV] & & [eV] & [eV] & & [nT]            \\  \hline
1 & 2022-02-25   & E11      &       320.4      &    461.3    &     1018.7  &   6.99          &  $6.8\times10^{-3}$    &       0.08      &  0.006    &    0.73      &     433 & 0.32 & 28.0        &   120.4      & 4.3        &      624.6       &  1016.9  & 1.63    &  643.9\\
 & 21:50:50UT    &       &               &     &        &           &   &            &     &        &   &  &    &     &      &       &   &     &   \\
2 &2022-02-25  & E11      &     315.8	&  497.5	   & 1096.8.     & 7.63            &  $6.9\times10^{-3}$    &       0.08     &  0.006	    &   0.72    &     433 & 0.42 & 31.85     &  107.04    & 3.4         &    503.8   & 1032.2 & 2.05	 & 667.4\\
 & 18:56:38 UT    &       &               &     &        &           &   &            &     &        &   &  &    &     &      &       &   &     &   \\
 3 & 2021-11-22   & E10      &      172.0         &   311.2    &   1439.4     &   19.3         &  $1.3\times10^{-2}$ &         0.23        &    0.033    &     0.78     &    223 & 0.62 & 135.9   &   26.33    & 0.19        &      749.5      &    640.16 & 0.85    & 397.9 \\
 & 02:42:10UT    &       &               &     &        &           &   &            &     &        &   &  &    &     &      &       &   &     &   \\
4 & 2022-06-03   & E12      &       224.5      &    302.6    &    799.25  &   5.56          &  $6.9\times10^{-3}$   &       0.10      &  0.007    &    0.91      &      248 & 0.32 & 21.17     &     39.37    & 1.86        &      353.8      &  355.6 &  1.0     &  325.9 \\
 & 04:17:01UT    &       &               &     &        &           &   &            &     &        &   &  &    &     &      &       &   &     &   

\end{tabular}
\label{pspdat:tab}
\end{sidewaystable}

%\end{table}

%\newpage
\section{Summary and Conclusions} \label{disc:sec}
Recently, PSP crossed the Alfv\'{e}nic surface and entered the sub-Alfv\'{e}nic solar wind close to the Sun for the first time near 16 $R_s$ \citep{Kas21}. Since then, PSP has traversed  the sub-Alfv\'{e}nic SW regions for increasing time periods during each encounter.  The solar wind is magnetically dominated in the sub-Alfv\'enic medium and its condition is of particular interest for understanding SW plasma instabilities, acceleration, and heating mechanisms. In particular, in these regions inward propagating Alfv\'{e}n wave packets can interact with outward propagating Alfv\'{e}n wave packets, leading to enhanced turbulence generation \citep[as originally proposed by][and since than extensively studied]{Kra65} and dissipation. We analyze several examples of sub-Alfv\'{e}nic solar wind data from PSP SPAN-I and FIELDS observations at perihelia encounters E10-E12, using the protons and $\alpha$ particle data, the constructed proton VDFs, the magnetic field, betas and temperature anisotropies, as well as kinetic wave properties. SPAN-I observations of the 3D VDFs demonstrate that PSP is capable of showing non-Maxwellian features of the young solar wind in the sub-Alfv\'{e}nic region. The observed non-Maxwellian (non-equilibrium) features in VDFs give rise to anisotropic temperatures and have free energies to drive waves and instabilities in the solar wind plasma. 

The solar wind plasma parameters and VDFs obtained from the PSP SPAN-I data analysis in the sub-Alfv\'{e}nic SW are used to initialize 2.5D and 3D hybrid models that study the evolution of proton and $\alpha$ particle populations and the kinetic instabilities driven by ion temperature anisotropies. The modeling results show the temporal evolution of the temperature anisotropies, the parallel and perpendicular kinetic energies, magnetic wave energy and heating, as well as the associated ion-scale wave dispersion in the SW plasma rest frame. We find that when proton temperature anisotropy is $A_p>2$ in the low-$\beta$ plasma, the proton VDF exceeds the ion-cyclotron instability threshold, resulting in the eventual relaxation of the temperature anisotropy through perpendicular proton cooling, parallel proton heating, and generation of ion-cyclotron wave spectrum. At the same time the $\alpha$ particles are strongly heated in the perpendicular direction through resonance with some of the proton emitted kinetic wave spectrum, with subsequent relaxation of the $\alpha$ population temperature anisotropy through a secondary ion-cyclotron instability. The models produce the evolution of the non-Maxwellian VDFs of $\alpha$  particles, and the isotropization of the proton VDF at later times. A self-consistent kinetic wave spectrum produced by an initially unstable proton VDF was modeled exhibiting the left-hand resonant (with ions) and right-hand nonresonant polarized dispersion branches. The spectra and dispersion of the ion-scale kinetic waves obtained  using the SW parameters of the various encounters demonstrate the increased wave-associated magnetic energy  and the evolution due to the ion-cyclotron instability. The left- and right-hand polarized  branches of the  dispersion relation obtained from the nonlinear hybrid model are in qualitative agreement with linear dispersion relations and with FIELDS observations of increased ion-scale kinetic wave activity associated with the unstable ion VDFs. Comparison of the 2.5D hybrid modeling results to more realistic but computationally intensive full 3D hybrid models shows very good agreement of the temporal evolution of the proton and $\alpha$ particle temperature anisotropies, and of the resulting ion VDFs. The good agreement between the 2.5D and 3D hybrid models suggests that the 3D spatial effects have small influence on the nonlinear evolution of the temperature anisotropy driven instabilities, that is dominated by parallel propagating modes.

The combined observational analysis of several time intervals at E10-E12, the constructed proton VDFs and the modeling results  demonstrate the importance of ion kinetic instabilities as the source of parallel heating of protons, the generation of ion-scale kinetic wave spectrum such as ion-cyclotron waves, and the resonant perpendicular as well as parallel heating of $\alpha$ particles. In particular, we find that the relaxation of proton kinetic instabilities in the sub-Alfv\'{e}nic regions of the solar wind can significantly contribute to the preferential heating and acceleration of $\alpha$ particle population. At present the source of the observed large proton perpendicular anisotropy as well as the generation of small scale fluctuations down to kinetic scales parallel to the magnetic field are not well understood. Our model demonstrates that the large temperature anisotropy observed at perihelia in the sub-Alfv\'{e}nic wind for several case studies leads to kinetic instability and relaxes rapidly  on timescale of $\sim$10-100 s. Thus, the observed unstable anisotropic proton and $\alpha$ particle VDFs  cannot be sustained for long time scales, without replenishing locally in the sub-Alfv\'{e}nic region of the solar wind by  processes such as large amplitude MHD waves, shocks, and fluid-scale turbulence that cascades to kinetic scales. 

%\subsection{Implication for PSP Observations}\label{psp_impact:sec}

\begin{acknowledgments}
The authors LJ and LO acknowledge support by NASA LWS grant 80NSSC20K0648. LO acknowledges support by NASA Goddard Space Flight Center through Cooperative Agreement 80NSSC21M0180 to Catholic University, Partnership for Heliophysics and Space Environment Research (PHaSER). PM acknowledges the support from NASA HGIO grant 80NSSC23K0419. JLV acknowledges support from NASA PSP-GI 80NSSC23K0208 and NASA LWS 80NSSC22K1014. 
Resources supporting this work were provided by the NASA High-End Computing (HEC) Program through the NASA Advanced Supercomputing (NAS) Division at Ames Research Center. This paper benefited from the discussions at International Space Science Institute (ISSI) in Bern, through ISSI International Team project 563 (Ion Kinetic Instabilities in the Solar Wind in Light of Parker Solar Probe and Solar Orbiter Observations).
%The authors thank Dr. Vadim Roytershtein for providing the linear Vlasov dispersion solver.
\end{acknowledgments}


%\vspace{5mm}
\facility{PSP (SWEAP, FIELDS)}

\end{document}